%% file: TensorOptVLDB.tex
\documentclass{vldb}
\usepackage{graphicx}
\usepackage{balance}  

\usepackage{microtype}
\usepackage{graphicx}
\usepackage{subcaption}
\usepackage{booktabs}
\usepackage{amsmath}

\usepackage{hyperref}
\usepackage{xcolor}
\usepackage{listings}

\usepackage{nicefrac}       

\usepackage{algorithm, algorithmic}
\usepackage{tabularx}

\definecolor{dkgreen}{rgb}{0,0.6,0}
\definecolor{gray}{rgb}{0.5,0.5,0.5}
\definecolor{mauve}{rgb}{0.58,0,0.82}

\newtheorem{linear dp complexity}{Theorem}
\newtheorem{optcnn complexity}[linear dp complexity]{Theorem}

\newtheorem{frontier complexity}{Lemma}
\newtheorem{random size}[frontier complexity]{Lemma}
\newtheorem{one step complexity}[frontier complexity]{Lemma}
\newtheorem{node elimination complexity}[frontier complexity]{Lemma}

\newtheorem{random tuple}{Assumption}

\newtheorem{cost frontier}{Definition}

\vldbTitle{TensorOpt: Exploring the Tradeoffs in Distributed DNN Training with Auto-Parallelism}
\vldbAuthors{Zhenkun Cai et al.}
\vldbDOI{https://doi.org/10.14778/xxxxxxx.xxxxxxx}
\vldbVolume{12}
\vldbNumber{xxx}
\vldbYear{2019}

\makeatletter
\def\@copyrightspace{\relax}
\makeatother
\begin{document}


\title{TensorOpt: Exploring the Tradeoffs in Distributed DNN Training with Auto-Parallelism}



%
%
%
%

\numberofauthors{1} 

\author{
  %
  %
  \alignauthor
  Zhenkun Cai, Kaihao Ma, Xiao Yan, Yidi Wu, Yuzhen Huang, James Cheng,\\ *Teng Su, *Fan Yu\\
  \affaddr{The Chinese University of Hong Kong, *Huawei Technologies Co. Ltd}\\
  \email{\{zkcai, khma, xyan, ydwu, yzhuang, jcheng\}@cse.cuhk.edu.hk\\ \{*suteng, *fan.yu\}@huawei.com}
}

\maketitle

\input{TO_abstract}

\input{TO_introduction}

\input{TO_related}
\input{TO_algorithm_new}
\input{TO_system}

\input{TO_analysis}

\input{TO_conclusion}

\bibliography{TensorOpt}
\bibliographystyle{plain}

\end{document}

%% file: TO_abstract.tex
\begin{abstract}
	

A good parallelization strategy can significantly improve the efficiency or reduce the cost for the distributed training of deep neural networks (DNNs). Recently, several methods have been proposed to find efficient parallelization strategies but they all optimize a single objective (e.g., execution time, memory consumption) and produce only one strategy. We propose \textit{FT}, an efficient algorithm that searches for an optimal set of parallelization strategies to allow the trade-off among different objectives. FT can adapt to different scenarios by minimizing the memory consumption when the number of devices is limited and fully utilize additional resources to reduce the execution time. For popular DNN models (e.g., vision, language), an in-depth analysis is conducted to understand the trade-offs among different objectives and their influence on the parallelization strategies. We also develop a user-friendly system, called \textit{TensorOpt}, which allows users to run their distributed DNN training jobs without caring the details of parallelization strategies. Experimental results show that FT runs efficiently and provides accurate estimation of runtime costs, and TensorOpt is more flexible in adapting to resource availability compared with existing frameworks.

       
\end{abstract}

%% file: TO_introduction.tex
\section{Introduction}\label{sec:intro}

Deep learning has been undergoing rapid developments in recent years~\cite{resnet,transformer,AmoebaNet} and the state-of-the-art deep neural networks (DNNs) are becoming increasingly difficult to train. Vision models~\cite{VeryDeepCNN} can take weeks to train on a single GPU~\cite{inception} and language models~\cite{googleNMT, nmt} can consume hundreds of gigabytes (GBs) of memory~\cite{gpipe}. The demands for intensive computation and large memory call for distributed training with multiple devices, which is typically conducted in GPU clusters~\cite{dean2012large,Tensorflow}.

A fundamental problem of distributed DNN training is finding a good parallelization strategy. A \textit{parallelization strategy} is a partitioning and assignment of the operators of a DNN model to devices (e.g., GPUs), and the strategy is associated with its runtime costs for training the DNN model, including execution time, memory consumption and network communication time\footnote{Execution time refers to the time taken for a mini-batch and memory consumption is the peak memory consumption for training a mini-batch, which includes both the model parameter and activations.}. Two simple and widely used parallelization strategies are \textit{data parallelism} and \textit{model parallelism}~\cite{dean2012large,kim2017splitnet}. Data parallelism keeps a copy of the entire model on each device and synchronizes the model copies in each mini-batch. Model parallelism assigns a disjoint set of layers to each device and communicates the activations across the devices. However, data parallelism is inefficient for layers with large parameters (e.g., fully connected layers) and model parallelism suffers from high communication cost when the activations are large (e.g., convolution layers).

Recently, more advanced methods~\cite{OptCNN,FlexFlow,Tofu} have been proposed to find parallelization strategies that are much more efficient than simple data parallelism and model parallelism. We call these methods \textit{auto-parallel} or \textit{auto-parallelism}. Their success lies in searching the large space of possible parallelization strategies using well-designed algorithms. OptCNN~\cite{OptCNN} minimizes the execution time for training convolutional neural networks (CNNs) with a dynamic programming algorithm. FlexFlow~\cite{FlexFlow} considers a more diverse set of DNN models, e.g., recurrent neural networks (RNNs)~\cite{lstm}, and minimizes the execution time using a randomized Markov Chain Monte Carlo (MCMC) search algorithm. ToFu~\cite{Tofu} focuses on training large models and minimizes the memory consumption. As training large models (those that can not be placed in memory of a single device) is becoming increasingly important~\cite{pipedream,gpipe}, TensorFlow provides the Mesh-TensorFlow library~\cite{MeshTensorFlow} to allow users to program their own parallelization strategies. 

Existing works only optimize a single objective (execution time or memory consumption), which results in \textit{limited flexibility to adapt to different scenarios}. For example, when training large models using a small number of devices, simply minimizing the execution time could result in memory overflow. We also found that methods that minimize the memory consumption cannot fully utilize additional memory resource to reduce the execution time. In some cases, \textit{it is important to track the trade-offs among different objectives}. For example, knowing the minimum execution time of a training job when using different amounts of resources (e.g., memory and number of devices) can help us make resource allocation decisions in a shared GPU cluster. When training DNNs on the cloud, users need to know the trade-offs between the cost (i.e., resources) and the efficiency (i,e., execution time) to determine the amount of resources to purchase. Therefore, the algorithm should be flexible enough to find parallelization strategies according to specific scenarios and user preferences (on the cost-efficiency trade-off), rather than optimizing a single objective. Moreover, an auto-parallel system should make finding and programming parallelization strategies transparent to users as both tasks require a deep understanding of distributed DNN training.

In this paper, we make three main contributions. First, we formulate the concept of \textit{cost frontier} and propose the \textit{Frontier-Tracking} (\textit{FT}) algorithm to find the cost frontier efficiently. For a given DNN model and device configuration, the cost frontier is a minimum set of parallelization strategies, $\mathcal{F}$, such that given any parallelization strategy $S$, there exists a strategy in $\mathcal{F}$ that gives a smaller or equal cost in every dimension (e.g., execution time, memory consumption) than $S$. Thus, parallelization strategies outside the cost frontier are not attractive as we can always find strategies in the frontier that outperform them. The cost frontier also provides a continuum for the trade-offs among different objectives and allows users to flexibly choose a parallelization strategy according to their scenario (e.g., resource availability in a cluster, cloud resource budget). As the complexity of finding the cost frontier by brute-force search is exponential (w.r.t. the number of operators in a given DNN model), the FT algorithm adopts a carefully designed dynamic programming procedure for efficient cost frontier tracking. Our analysis shows that the complexity of the FT algorithm is only quadratic in the number of operators in a given DNN model.  

Second, we propose a flexible and user-friendly auto-parallel system called \textit{TensorOpt}. TensorOpt uses TensorFlow as the underlying execution engine and its API is almost identical to TensorFlow, so that users only need to make a few changes to run their TensorFlow scripts as auto-parallel jobs on TensorOpt. TensorOpt also makes parallelization strategy search and implementation totally transparent to users by using the FT algorithm for strategy search and generating the low-level execution graph automatically according to the chosen parallelization strategy. Users only need to specify their preference for the parallelization strategy via some high-level options. By removing the tensor split restrictions in MeshTensorFlow~\cite{MeshTensorFlow}, TensorOpt allows a larger space for parallelization strategy search and hence better performance.

Third, we conducted extensive experiments to characterize the cost frontier and validate the effectiveness of the FT algorithm and the TensorOpt system. For all the models we experimented, we found that there exists a sharp turning point in the trade-off between memory consumption and execution time. The execution time increases rapidly when available memory is below the turning point but drops slowly when more memory is provided. We also found that both inter-machine and intra-machine communication bandwidth play a decisive role in the efficiency of distributed DNN training. Thanks to the FT algorithm, TensorOpt is flexible in adapting to different scenarios, i.e., TensorOpt can choose strategies to minimize memory consumption when the number of device is limited and fully utilize additional resources to minimize execution time. Moreover, both the FT algorithm and TensorOpt system have good efficiency.

%% file: TO_related.tex
\section{Background and Related Work}\label{sec:related}

We first provide some background. Then we discuss related work and their limitations, which motivate our work.

\subsection{Parallelization Strategy and Execution Cost}\label{subsec:background}

\begin{figure*}[!t]
  \centering
  \includegraphics[width=1\columnwidth]{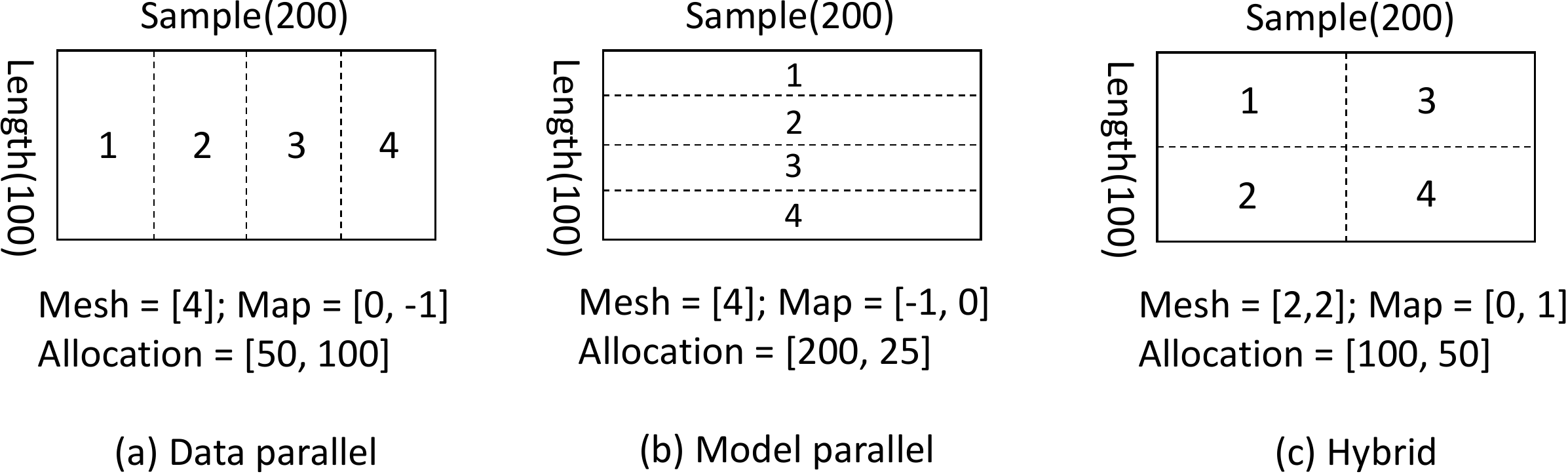}
  \vspace{-1mm}
  \caption{Examples of simple parallelization configurations for the input tensor of an operator that computes matrix-vector product with the matrix being the model parameter. The size of the input tensor is [200, 100] with 200 being the batch size and 100 being the length of the vector. There are 4 GPUs, which are represented by numbers 1-4.}
  \label{fig:demo of parallelization configurations}
  \vspace{-3mm}
\end{figure*}


We first define the notations used in this paper. The computation devices (e.g., GPUs) are modeled as a device graph $\mathcal{D}$, with each node $d_i$ being a device and each edge $(d_i, d_j)$ being the network connection between $d_i$ and $d_j$. A DNN is modeled as a computation graph $\mathcal{G}$, in which nodes are operators and a directed edge $e_{ij}$ means that the output tensor of operator $o_i$ is used as the input for operator $o_j$. We focus on synchronous training, although our method can also be extended to asynchronous training (e.g., as in PipeDream~\cite{pipedream}) by changing the cost functions.


\vspace{1mm}
\textbf{Parallelization configurations.} A parallelization strategy $S$ contains a parallelization configuration $s_i^k$ for each operator $o_i$ in the computation graph $\mathcal{G}$ and determines how the devices execute the training job. $s_i^k$ is selected from a set $\mathcal{S}_i$ that contains $K_i$ valid parallelization configurations for $o_i$, where $1 \le k \le K_i$. More specifically, a parallelization configuration $s_i^k$ consists of a \textit{device mesh} and some \textit{tensor maps}, which jointly describe how the tensors (both input and model parameter) related to an operator are split among the devices. Following MeshTensorFlow~\cite{MeshTensorFlow}, the device mesh is an integer array used to describe the logical organization of the devices. For example, 4 GPUs can be represented as [4] (as a one-dimensional array) or [2, 2] (as a two-dimensional array). A tensor map is an integer array with its size being the dimension of the tensor and describes how each dimension of the tensor is split on the device mesh. Consider an operator that computes matrix-vector product (with the matrix being the model parameter) with an input size of [200, 100], where 200 is the batch size and 100 is the vector length. With a device mesh [2, 2], a tensor map of [0, 1] for the input tensor means that the first dimension of the input is split across the first dimension of the device mesh, and the second dimension of the tensor is split across the second dimension of the mesh. As a result, each device will have a slice of the input tensor with shape [100, 50]. If -1 is used in the tensor map, then the corresponding tensor dimension is not split across any mesh dimension. More examples of parallelization configurations are shown in Figure~\ref{fig:demo of parallelization configurations}. We have developed a complete set of rules to decide what are the valid parallelization configurations for an operator (e.g., redundant computation of the same tensor on different devices is also allowed for possible memory/communication saving). The details will be released together with the code (we will open source TensorOpt) and are omitted here for conciseness. As $\mathcal{S}_i$ contains all feasible combinations of the device mesh and tensor maps, it can be very large when the number of devices and/or the dimension of the tensors is large.

\vspace{1mm}
\textbf{Execution cost.} For operator $o_i$ under parallelization configuration $s_i^k$, its memory cost and time cost $(m(o_i, s_i^k),$ $t(o_i, s_i^k))$ are defined as follows
\begin{equation}\label{equ:operator cost}
  \begin{aligned}
    m(o_i, s_i^k) & =m_p(o_i, s_i^k)+m_t(o_i, s_i^k), \\
    t(o_i, s_i^k) & =t_c(o_i, s_i^k)+t_s(o_i, s_i^k),
  \end{aligned}
\end{equation}
where $m_p(o_i, s_i^k)$ is the memory for storing the (partitioned) model parameter, $m_t(o_i, s_i^k)$ is the memory for storing temporary tensors (e.g., tensors for use in backward propagation)\footnote{There are some other memory consumptions, e.g., for kernel execution and network communication, but we found that these consumptions are relatively much smaller.}, $t_c(o_i, s_i^k)$ is the time taken to conduct the computation defined by operator $o_i$ (including both forward pass and backward pass), and $t_s(o_i, s_i^k)$ is the time taken to synchronize the tensors associated with $o_i$ (e.g., for model parameter update in data parallel). Among them, $m_p(o_i, s_i^k)$ and $m_t(o_i, s_i^k)$ can be derived from the specification of $o_i$ in $\mathcal{G}$ and the parallelization configuration $s_i^k$, while $t_c(o_i, s_i^k)$ and $t_s(o_i, s_i^k)$ are measured by running the operator under the parallelization configuration multiple times. We also call the memory cost and time cost in Eq.~\eqref{equ:operator cost} the \textit{operator costs}.

For edge $e_{ij}$, its memory cost and time cost $(m(e_{ij}, s_i^k, s_j^p),$ $t(e_{ij}, s_i^k, s_j^p))$ are defined as
\begin{equation}\label{equ:edge cost}
  \begin{aligned}
    m(e_{ij}, s_i^k, s_j^p) & =0,                         \\
    t(e_{ij}, s_i^k, s_j^p) & =t_x(e_{ij}, s_i^k, s_j^p),
  \end{aligned}
\end{equation}
where $t_x(e_{ij}, s_i^k, s_j^p)$ is the time taken to transfer the tensors between operator $o_i$ and operator $o_j$ (including both forward pass and backward pass), which depends on the parallelization configuration of both $o_i$ and $o_j$ (i.e., $s_i^k$ and $s_j^p$). We call the costs in Eq.~\eqref{equ:edge cost} the \textit{edge costs}.

With the costs of each individual operator and edge, we can define the execution time (or per-iteration time) $t(S, \mathcal{G}, \mathcal{D})$, peak memory consumption $m(S, \mathcal{G}, \mathcal{D})$, and communication cost $c(S, \mathcal{G}, \mathcal{D})$ of a complete parallelization strategy $S$ for the entire computation graph $\mathcal{G}$ as follows
\begin{equation}\label{equ:total cost}
  \begin{aligned}
    t(S, \mathcal{G}, \mathcal{D}) & =\sum_{\substack{o_i\in \mathcal{G} \\ s_i^k \in S}} t(o_i, s_i^k)+\sum_{\substack{e_{ij}\in \mathcal{G}\\s_i^k \in S, s_j^p \in S}} t(e_{ij}, s_i^k, s_j^p),\\
    m(S, \mathcal{G}, \mathcal{D}) & =\sum_{\substack{o_i\in \mathcal{G} \\ s_i^k \in S}} m(o_i, s_i^k),\\
    c(S, \mathcal{G}, \mathcal{D}) & =\sum_{\substack{o_i\in \mathcal{G} \\ s_i^k \in S}} t_s(o_i, s_i^k)+\sum_{\substack{e_{ij}\in \mathcal{G}\\s_i^k \in S, s_j^p \in S}} t_x(e_{ij}, s_i^k, s_j^p).
  \end{aligned}
\end{equation}


\subsection{Related Work}\label{subsec:existing work}

\textbf{Data and model parallelism.} Data parallelism~\cite{data_parallel} is a common parallelization strategy adopted by deep learning frameworks including TensorFlow~\cite{Tensorflow}, PyTorch~\cite{PyTorch} and MxNet~\cite{mxnet}. It keeps a copy of the model on each device and partitions the input tensor among the devices along the sample (batch) dimension. Compared with data parallelism, model parallelism~\cite{dean2012large, RLDevicePlacement} is more suitable for large models, e.g., those that do not fit in the memory of a single device, as it partitions the model among the devices to alleviate the memory consumption problem. However, the resource utilization of vanilla model parallelism is low as the devices execute different partitions of the model sequentially. Due to the increasing interest in training large models, recent works improve model parallelism with \textit{pipeline parallelism}. Gpipe~\cite{gpipe} splits a mini-batch into several micro-batches and pipelines these micro-batches to reduce device idle time. PipeDream~\cite{pipedream} removes the mini-batch synchronization barrier in Gpipe to further improve device utilization but training becomes asynchronous. A dynamic programming algorithm is also proposed in PipeDream to find the model partitioning that minimizes the per-iteration time. However, asynchronous training often degrades the convergence speed of training and some models even cannot converge~\cite{ssp}.

%
%
%
%
%

\vspace{1mm}
\textbf{Manual strategies.} It has long been observed that pure data or model parallelism may not achieve the best performance. One-wired-trick~\cite{OneWierdTrick} manually designs a parallelization strategy for CNN, which uses data parallelism for the convolution layers and model parallelism for the fully connected layers. Mesh-TensorFlow~\cite{MeshTensorFlow} provides a flexible parallel training framework that allows users to specify their parallelization strategies. However, users need to find a good parallelization strategy by themselves and manually program it in the code, both of which requires a good understanding of parallel training. Moreover, Mesh-TensorFlow has some restrictions on the parallelization strategies and we will show that these restrictions lead to sub-optimal performance.

%

\vspace{1mm}
\textbf{Auto-parallel.} Recently, some works propose to search for efficient parallelization strategies for DNN training using tailored algorithms. OptCNN~\cite{OptCNN} uses dynamic programming (DP) to find the parallelization strategy that minimizes the per-iteration time. The DP algorithm simplifies the  model computation graph into a graph that contains only two nodes by conducting node and edge elimination, and finds the optimal strategy on the simplified graph using brute-force search. As OptCNN considers only execution time, its parallelization strategy may go out of memory for large models or when memory is limited. Moreover, the node and edge elimination of OptCNN is not sufficient for some models (e.g., BERT~\cite{bert}). FlexFlow~\cite{FlexFlow} works for a wider range of models using a random search algorithm to find the parallelization strategy. However, FlexFlow also only considers execution time and the parallelization strategy it produces may not be optimal. ToFu~\cite{Tofu} minimizes the memory consumption for training large models using DP. The DP algorithm splits a tensor among two (groups of) devices each time to reduce complexity and ToFu does not allow tensor replication to achieve low memory consumption. However, ToFu cannot leverage additional memory (those more than the minimum requirement) to reduce the execution time.

%
%

\vspace{1mm}
\textbf{Memory optimizations.} Some works reduce the memory consumption for training large models with extra communication or computation costs~\cite{vdnn, chen2016training, memoryopt}. VDNN~\cite{vdnn} swaps
the tensors from GPU to CPU and reloads them for backward propagation to reduce peak memory consumption. ~\cite{chen2016training} only keeps some of the tensors in memory and recomputes the other tensors when needed in backward propagation. However, these extra communication or computation costs may significantly degrade the training performance. Our methods could be extended by considering reloading and re-computation as possible parallelization configurations.

Compared with the related works, our FT algorithm and TensorOpt system significantly improve both \textit{flexibility} and \textit{usability}. By tracking the cost frontier, FT can adapt to different scenarios, e.g., reducing the memory consumption when the model is large and/or memory is limited, while minimizing the execution time when memory is sufficient. FT can also fully utilize available resources and translate additional resources (e.g., memory) into performance improvements. Compared with Mesh-TensorFlow, TensorOpt is much more user-friendly by using the FT algorithm to search for the parallelization strategy and automatically executing the parallelization strategy. Users only need to define the computation graph using the high-level API (as in vanilla TensorFlow) and specify their preferences for the parallelization strategy.

%
%
%

%% file: TO_algorithm_new.tex
\section{The Frontier-Tracking Algorithm}\label{sec:alg}

In this section, we first introduce the concept of cost frontier. As using brute-force search to find parallelization strategies on the cost frontier has very high complexity, we propose an efficient frontier-tracking (FT) algorithm. Finally, we conduct analysis to validate the low complexity of the FT algorithm. For simplicity, we present the cost frontier and the FT algorithm for tracking the trade-off between execution time $t(S, \mathcal{G}, \mathcal{D})$ and memory consumption $m(S, \mathcal{G}, \mathcal{D})$, while generalizing our methods to tracking the trade-off between any pair of costs (e.g., memory consumption and network communication) should be straightforward.

\subsection{Cost Frontier}\label{subsec:cost frontier}

\begin{figure}[!t]
	\centering
	\includegraphics[width=0.50\columnwidth]{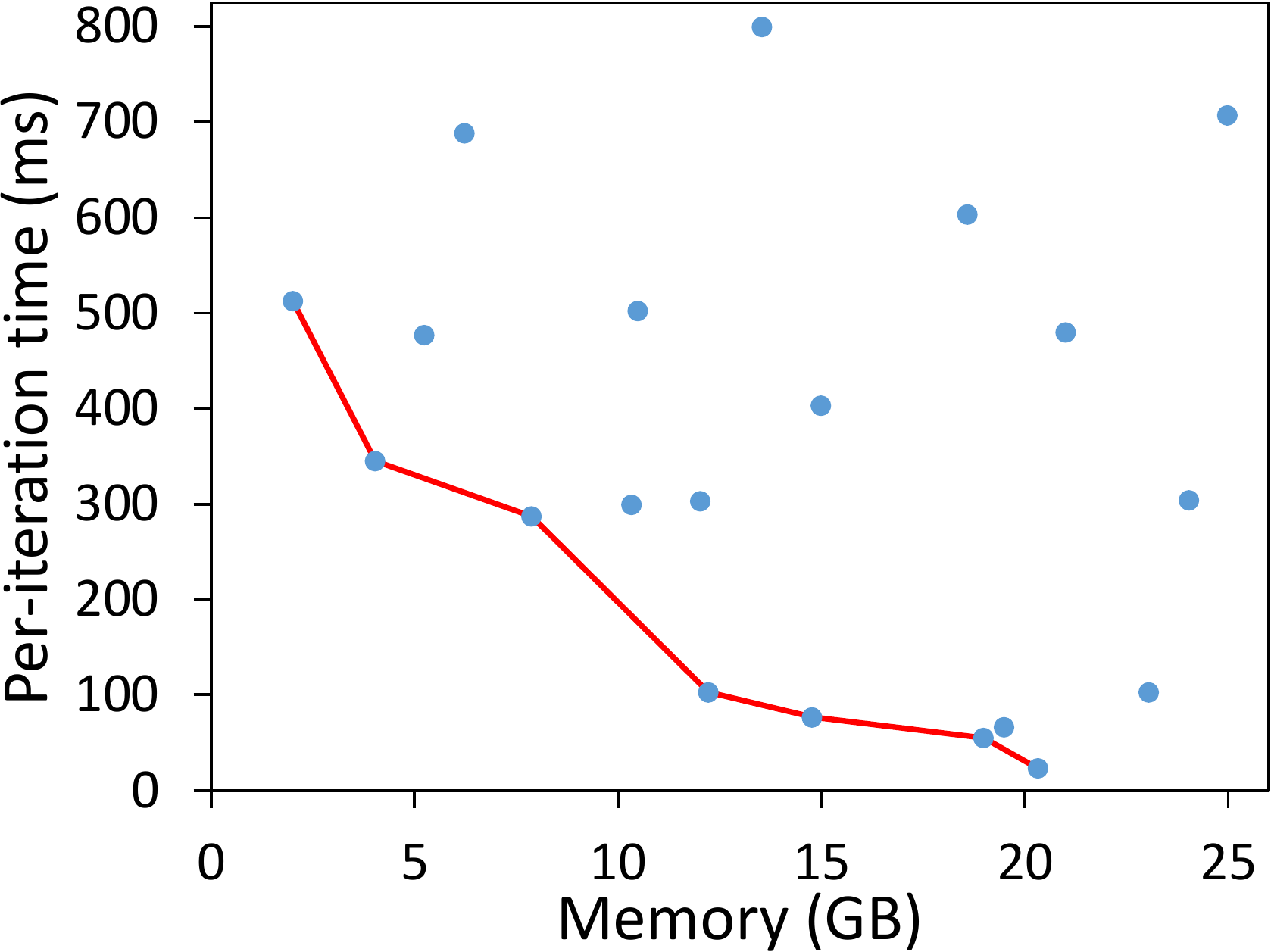}
	\vspace{-2mm}
	\caption{An illustration of cost frontier}
	\label{fig:cost frontier}
	\vspace{-2mm}
\end{figure}

The formal definition of \textit{cost frontier}  is given as follows. 
\begin{cost frontier}\label{def:frontier}        
Let $\mathcal{C}\!=\!\{(S_1, m_1, t_1), (S_2, m_2, t_2),\cdots, (S_K,\\ m_K, t_K)\}$ be a set of (partial) parallelization strategy tuples, where $S_k$ is a (partial) parallelization strategy, $m_k$ and $t_k$ are the execution time and memory consumption of $S_k$,  for $1\le k \le K$. The cost frontier of $\mathcal{C}$ is the minimum subset $\mathcal{F}$ of $\mathcal{C}$ such that, for any strategy $(S_p, m_p, t_p)\in \mathcal{C}$, there exists a strategy  $(S_k, m_k, t_k)\in \mathcal{F}$ where $m_k \leq m_p$ and $t_k \leq t_p$.
\end{cost frontier}
We provide an illustration of cost frontier in Figure~\ref{fig:cost frontier}, in which each point is a strategy tuple with randomly generated costs and points on the line are the cost frontier. According to Definition~\ref{def:frontier}, for a strategy that is not in the cost frontier, we can find some strategy in the frontier that can reduce at least one of the two costs without increasing the other. Therefore, it suffices to find all parallelization strategies in the frontier of $t(S, \mathcal{G}, \mathcal{D})$ and $m(S, \mathcal{G}, \mathcal{D})$. Users can choose a parallelization strategy in the frontier according to their situation. For example, if memory is sufficient, the parallelization strategy that minimizes per-iteration time can be used. When memory is limited, users can choose the strategy that minimizes memory consumption instead.          

Given a set $\mathcal{C}$ of strategy tuples, its cost frontier can be obtained using Algorithm~\ref{alg:frontier}. In Algorithm~\ref{alg:frontier}, $\mathcal{C}_m$ is the list obtained by sorting $\mathcal{C}$ in ascending order of memory consumption, $\mathcal{C}_m[i]$ is the $i^{\textsubscript{th}}$ tuple in $\mathcal{C}_m$, and $t(\mathcal{C}_m[i])$ denotes the time consumption of tuple $\mathcal{C}_m[i]$. Algorithm~\ref{alg:frontier} checks the tuples in ascending order of memory consumption and puts a tuple into $\mathcal{F}$ if it has smaller time consumption than all tuples that precede it in $\mathcal{C}_m$. In the $i\textsuperscript{th}$ step, $v$ records the smallest time consumption from $\mathcal{C}_m[1]$ to $\mathcal{C}_m[i-1]$.

\begin{algorithm}
  \caption{Reduce to frontier}
  \label{alg:frontier}
  {\small
  \begin{algorithmic}[1]
    \STATE {\bfseries Input:} A set $\mathcal{C}$ containing $K$ strategy tuples
    \STATE {\bfseries Output:} The cost frontier $\mathcal{F}$ of $\mathcal{C}$
    \STATE Sort tuples in $\mathcal{C}$ in ascending order of memory consumption and denote the result as $\mathcal{C}_m$
    \STATE Initialize $\mathcal{F}=\emptyset$, $v=+\infty$
    \FOR{\texttt{$i = 1$ to $K$}}
    \IF{$t(\mathcal{C}_m[i])<v$}
    \STATE $\mathcal{F}=\mathcal{F}\cup \mathcal{C}_m[i]$ and $v=t(\mathcal{C}_m[i])$
    \ENDIF
    \ENDFOR
    \STATE Return $\mathcal{F}$
  \end{algorithmic}
}
\end{algorithm}

A straightforward method to track the cost frontier is to enumerate all possible parallelization strategies for the computation graph $\mathcal{G}$, calculate their memory and time consumption according to Eq.~\eqref{equ:total cost} and find the cost frontier by applying Algorithm~\ref{alg:frontier}. However, this method has an exponential complexity. Assume that $\mathcal{G}$ contains $n$ operators and each operator has $K$ parallelization configurations, this brute-force search needs to go through all $K^n$ parallelization strategies. As $\mathcal{G}$ usually contains tens or even hundreds of operators for popular DNN models, brute-force search is infeasible. Therefore,  we propose the FT algorithm to find the cost frontier efficiently given $\mathcal{G}$.  FT relies on the following~\textit{basic operations} to manipulate cost frontiers.               

 Given two cost frontiers (or two sets of strategy tuples),
\[
  \begin{aligned}
    \mathcal{F}  & =\{(S_1, m_1, t_1), (S_2, m_2, t_2),...,(S_K, m_K,t_K)\},                   \\
    \mathcal{F}' & =\{(S_1', m_1', t_1'), (S_2', m_2', t_2'),...,(S_{K'}', m'_{K'},t'_{K'})\}.
  \end{aligned}
\]

\begin{description}
  \item[$\bullet$]\textbf{Product}, which is the Cartesian product of two frontiers:
        $$\mathcal{F} \otimes \mathcal{F}' = \cup_{1\le k \le K, 1\le p \le K'}\left\{\big([S_k, S_p'], m_k+m_p', t_k+t_p'\big)\right\}.$$
  \item[$\bullet$]\textbf{Union}, which is the union of two frontiers:
        $$ \mathcal{F} \cup \mathcal{F}' = \cup_{1\le k \le K}\{(S_k, m_k,t_k)\}\,\,\cup \,\, \cup_{1\le p \le K'}\{(S_p', m_p',t_p')\}.$$
  \item[$\bullet$]\textbf{Reduce}, which is Algorithm~\ref{alg:frontier}, i.e., $\mathcal{F}=\text{reduce} (\mathcal{C})$. As the result of \textit{product} and \textit{union} may no longer be a frontier, we assume that \textit{reduce} is always applied after the two operations.
\end{description}

Intuitively, \textit{product} constructs composite parallelization strategies by enumerating all possible combinations of $S$ and $S'$, and the costs of $S$ and $S'$ are summed up in the product. The operation \textit{union} places the tuples from $\mathcal{F}$ and $\mathcal{F}'$ into a single set.

\vspace{-3mm}
\subsection{Frontier Tracking}\label{subsec:algorithm description}

\begin{figure*}[!t]
	\centering
	\includegraphics[width=1.35\columnwidth]{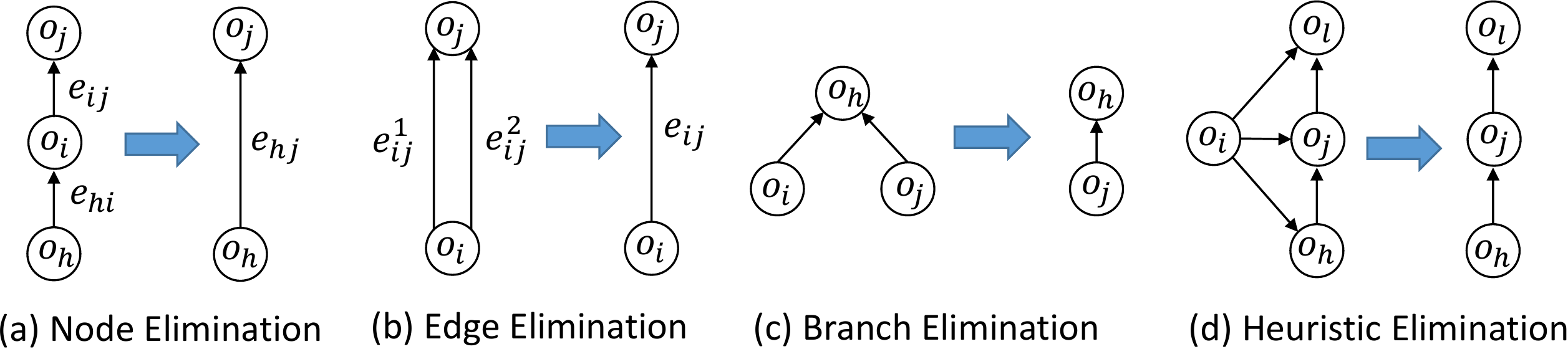}
	\vspace{-1mm}
	\caption{An illustration of different types of elimination in FT}
	\label{fig:elimination}
	\vspace{-2mm}
\end{figure*}  

\begin{figure}[!t]
	\centering
	\includegraphics[width=0.55\columnwidth]{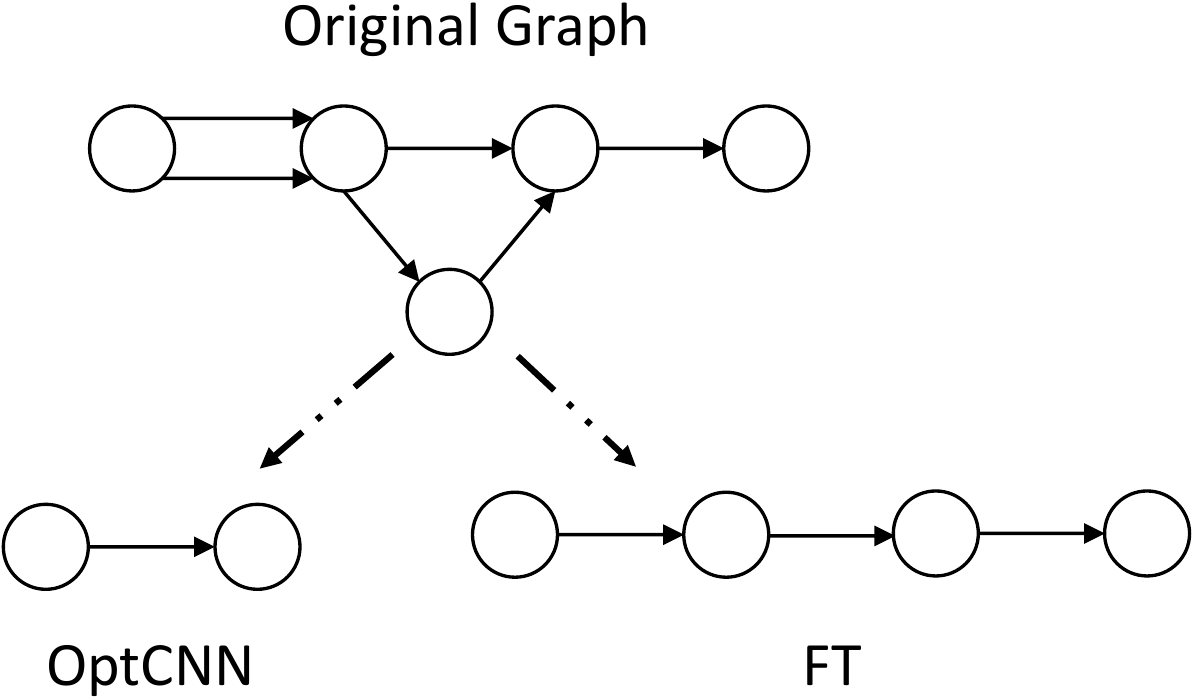}
	\vspace{-1mm}
	\caption{Difference between FT and OptCNN}
	\label{fig:linear graph}
	\vspace{-2mm}
\end{figure}

\textbf{Overview.} The procedure of our FT algorithm is shown in Algorithm~\ref{alg:FT}, which finds all parallelization strategies in the cost frontier of execution time and memory consumption for a given computation graph $\mathcal{G}$ and device graph $\mathcal{D}$. Algorithm~\ref{alg:FT} can be decomposed in 4 steps, i.e., \textit{initialization, elimination, linear dynamic programming (LDP) and unroll}. Initialization (Line~3) sets the cost frontier for each operator and edge in the computation graph $\mathcal{G}$. Elimination (Lines~4-11) simplifies the graph into a linear graph $\mathcal{G}'$ (as illustrated in Figure~\ref{fig:linear graph}) and updates the cost frontiers of the operators and edges. LDP (Line~12) finds the cost frontier for the simplified graph $\mathcal{G}'$ and unroll (Lines~13-14) reconstructs the parallelization strategies in the cost frontier for the original computation graph $\mathcal{G}$. In the following, we explain each of the 4 steps in more detail.      

\begin{algorithm} [!t]
	\caption{Frontier Tracking (FT)}
	\label{alg:FT}
	  {\small
	\begin{algorithmic}[1]
		\STATE {\bfseries Input:} Computation graph $\mathcal{G}$, device graph $\mathcal{D}$
		\STATE {\bfseries Output:} All parallelization strategies in the cost frontier of execution time and memory consumption
		\STATE Initialize all valid parallelization configurations and their costs for the operators and edges in $\mathcal{G}$
		\WHILE{true}
		\STATE Mark nodes on the linear graph
		\IF {not TryExactEliminate($\mathcal{G}$)}
		\IF {not TryHeuristicEliminate($\mathcal{G}$)}
		\STATE break
		\ENDIF
		\ENDIF
		\ENDWHILE
		\STATE Apply LDP in Algorithm~\ref{alg:Linear-dp} on the simplified graph $\mathcal{G'}$ and generate strategies on the cost frontier
		\STATE Unroll the LDP
		\STATE Unroll the elimination
	\end{algorithmic}
}
\end{algorithm}

\vspace{1mm}
\textbf{Initialization.} FT begins by initializing the costs for the edges and operators by enumerating all their possible parallelization configurations. With a slight abuse of the notations, we use $\mathcal{F}(o_i,s_i^k)$ to denote the tuple $(s_i^k, m(o_i, s_i^k), t(o_i, s_i^k))$ (i.e., the operator costs in Eq.~\eqref{equ:operator cost}), which is the cost frontier for operator $o_i$ when it selects $s_i^k$ as the parallelization configuration. Similarly, $\mathcal{F}(e_{ij},s_i^k,s_j^p)$ denotes the tuple $([s_i^k,s_j^p], m(e_{ij},s_i^k,s_j^p),t(e_{ij},s_i^k,s_j^p))$ (i.e., the edge costs in Eq.~\eqref{equ:edge cost}), which is the cost frontier for edge $e_{ij}$ when operator $o_i$ and $o_j$ use parallelization configuration $s_i^k$ and $s_j^p$. Although both $\mathcal{F}(o_i,s_i^k)$ and $\mathcal{F}(e_{ij},s_i^k,s_j^p)$ only have a cardinality of 1 when first initialized, their sizes may change when the FT algorithm updates them in elimination and LDP.

\vspace{1mm}
\textbf{Elimination.} FT conducts four types of elimination: \textit{node elimination}, \textit{edge elimination}, \textit{branch elimination}, and \textit{heuristic elimination}, to simplify the computation graph $\mathcal{G}$ into a linear graph $\mathcal{G}'$. The first three preserve the exact cost frontier, while heuristic elimination significantly reduces the complexity with only a small loss in accuracy. Compared with the two types of elimination (i.e., node and edge elimination) in OptCNN~\cite{OptCNN}, more types of elimination enable FT to adapt to a more diverse set of DNN models (e.g., BERT).  Moreover, for each type of elimination, FT maintains the cost frontier instead of a single execution time. We illustrate the four eliminations in Figure~\ref{fig:elimination} and discuss them as follows.

\vspace{1mm}

\textit{Node Elimination}. FT conducts node elimination when an operator has only one input operator and one output operator. As shown in Figure~\ref{fig:elimination}(a), $e_{hi}$, $o_i$ and $e_{ij}$ are replaced by a single edge $e_{hj}$ in node elimination. The cost frontier of $e_{hj}$ is deduced as follows
\begin{equation}\label{equ:node}
\begin{aligned}
\mathcal{F}(e_{hj}, s_h^w, s_j^p) & = \cup_{s_i^k\in\mathcal{S}_i}\{\mathcal{F}(e_{hi}, s_h^w, s_i^k) \otimes \mathcal{F}(o_i, s_i^k) \\
& \quad \quad \quad \quad \otimes \mathcal{F}(e_{ij}, s_i^k, s_j^p)\}.
\end{aligned}
\end{equation}
Under each combination of the parallelization configurations of operators $o_h$ and $o_j$, $o_i$ is eliminated by summing its operator cost to the costs on edge $e_{hi}$ and $e_{ij}$. Note that we apply~\textit{reduce} to the result of Eq.~\eqref{equ:node} to ensure that $\mathcal{F}(e_{hj}, s_h^w, s_j^p)$ is a frontier, which reduces the size of $\mathcal{F}(e_{hj}, s_h^w, s_j^p)$ and the complexity of subsequent operations. For each tuple in the frontier $\mathcal{F}(e_{hj}, s_h^w, s_j^p)$, FT records which parallelization configuration $o_i$ (i.e., $s_i^k$) takes to produce it in order to provide information for unrolling the elimination. 


\vspace{1mm}
\textit{Edge Elimination}. Edge elimination is conducted when there are multiple edges connecting the same pair of operators. Denote the edges as $(e_{ij}^1, e_{ij}^2, ...,e_{ij}^V)$, these edges are merged into a single edge $e_{ij}$, as illustrated in Figure~\ref{fig:elimination}(b). The cost frontier of the new edge is calculated as follows
\begin{equation}
\mathcal{F}(e_{ij}, s_i^k, s_j^p) = \otimes_{1\leq v \le V} \mathcal{F}(e_{ij}^v, s_i^k, s_j^p).
\end{equation}
Under the same parallelization configuration of the up-stream operator $o_i$ and down-stream operator $o_j$, the costs of the merged edges are added together for edge elimination. As node and edge elimination cannot simplify some complex computation graphs (e.g., BERT) to simple structures, we introduce \textit{branch elimination} and \textit{heuristic elimination}.  

\vspace{1mm}
\textit{Branch Elimination}. FT conducts branch elimination when an operator has multiple input operators and these operators cannot be eliminated by node or edge elimination. As shown in Figure~\ref{fig:elimination}(c), operator $o_h$ receives inputs from operators $o_i$ and $o_j$, and $o_i$ and $o_j$ cannot be eliminated because they are not connected by an edge. Branch elimination removes either $o_i$ or $o_j$ by merging it into $o_h$. If $o_i$ is merged, the cost frontier of $o_h$ is updated as follows
\begin{equation}
\begin{aligned}
\mathcal{F}(o_h, \tilde{s}_h^w) = & \cup_{s_i^k\in\mathcal{S}_i}\{\mathcal{F}(o_h, s_h^p) \otimes \mathcal{F}(o_i, s_i^k) \\
& \otimes \mathcal{F}(e_{ih}, s_i^k, s_h^p)\},
\end{aligned}
\end{equation}
where $\tilde{s}_h^w=(s_h^p,s_i^k)$ is the concatenation of the parallelization configuration of $o_i$ and $o_h$, and the costs of operator $o_i$ and edge $e_{ih}$ are added to the cost of operator $o_h$.

\vspace{1mm}
\textit{Heuristic Elimination}. FT conducts heuristic elimination when the three types of elimination introduced before cannot be applied. For example, the attention mask is used by all the transformer layers in BERT~\cite{bert} and thus cannot be eliminated. An illustration is shown in Figure~\ref{fig:elimination} (d), in which the computation graph cannot be simplified with other types of elimination. In this case, heuristic elimination simply decides the parallelization configuration for an operator $o_i$, and removes $o_i$ along with all its out-going edges. We use multiple heuristics to choose a parallelization configuration for $o_i$, e.g., minimizing the memory consumption of $o_i$ or a weighted combination of different objectives. After removing $o_i$ by selecting parallelization configuration $s_i^k$, an operator $o_j$ that takes input from $o_i$ updates its frontier as follows
\begin{equation}
\mathcal{F}(o_j, s_j^p) = \mathcal{F}(o_j, s_j^p) \otimes \mathcal{F}(e_{ij}, s_i^k, s_j^p),
\end{equation}
which adds the cost of edge $e_{ij}$ to operator $o_j$. Note that heuristic elimination does not guarantee to preserve the cost frontier. However, we found that it significantly reduces the running time of FT with only marginal loss in accuracy. This is because we usually conduct heuristic elimination for only a very small number of times. For example, heuristic elimination only needs to be used twice for BERT.



\vspace{1mm}
\textbf{LDP.} One can apply the aforementioned 4 types of eliminations to simplify the computation graph $\mathcal{G}$ into a graph that contains only two nodes and then find the cost frontier for the simplified graph by brute-force search. This method is similar to the algorithm in OptCNN~\cite{OptCNN} and we call it \textit{FT-Elimination}. However, we found that if the computation graph $\mathcal{G}$ has a linear structure (as shown in Figure~\ref{fig:linear graph}), its cost frontier can be found much more efficiently than conducting eliminations. Moreover, popular DNN models can be easily organized into a linear structure. For example, if we treat each residual block as a group for ResNet~\cite{resnet}, then the groups form a linear structure. For BERT~\cite{bert}, each transformer block can also be regarded as a group and the transformers form a linear structure.

Therefore, FT conducts elimination such that the resultant graph $\mathcal{G}'$ has a linear structure. We use a simple heuristic for this purpose in Algorithm~\ref{alg:FT} when choosing the nodes and edges to eliminate. Before elimination starts, we mark the first operator\footnote{According to topological order, ties are broken randomly.} in the computation graph $\mathcal{G}$. During elimination, we do not eliminate the marked operators, and checks if the last operator we marked has only one downstream operator. If so, we mark that downstream operator as it is also on a linear structure. After obtaining a linear graph, Algorithm~\ref{alg:Linear-dp} (LDP) is used to compute the cost frontier.

For Algorithm~\ref{alg:Linear-dp}, we assume that the cost frontiers of the operators and edges (i.e., $\mathcal{F}(o_i,s_i^k)$ and $\mathcal{F}(e_{ij},s_i^k,s_j^p)$) in the linear graph $\mathcal{G}'$ are properly initialized by the elimination procedures. The algorithm computes the cost frontier of $\mathcal{G}'$ from the operator that receives the initial input (numbered as $o_1$) to the operator that generates the final model output (numbered as $o_n$). For the first operator $o_1$, we initialize its \textit{cumulative frontier} $\mathcal{CF}(o_1, s^k_1)$ as $\mathcal{F}(o_1,s^k_1)$. For the $i^{th}$ operator, we use the product of $\mathcal{CF}(o_{i-1}, s^k_{i-1})$, the frontier $\mathcal{F}(e_{(i-1)i},s_{i-1}^k,s_i^p)$ of edge $e_{(i-1)i}$, and the operator frontier $\mathcal{F}(o_i, s_i^p)$ to derive $\mathcal{CF}(o_i, s_i^p)$. As a result, $\mathcal{CF}(o_i, s_i^p)$ represents the cumulative cost frontier from operator $o_1$ to $o_i$ when $o_i$ chooses parallelization configuration $s_i^p$. We only need to consider the partial strategy tuples (containing parallelization configurations from $o_1$ to $o_i$) in $\mathcal{CF}(o_i, s_i^p)$ when choosing the parallelization configuration for operator $o_{i+1}$. This is because for a tuple (denote as $S_i^k$) that does not belong to $\mathcal{CF}(o_i, s_i^p)$, there is at least one tuple in $\mathcal{CF}(o_i, s_i^p)$ (denote as $S_i^p$) that has lower time and memory consumption. As a result, $S_i^k$ cannot be in the cost frontier when we add the costs of operator $o_{i+1}$ and edge $e_{i(i+1)}$, which are common for both $S_i^k$ and $S_i^p$. Finally, LDP reduces the cumulative frontier at the last operator (i.e., $\cup_{s_n^k\in\mathcal{S}_n}\mathcal{CF}(n,s_n^k)$) to find the cost frontier for the entire graph (Line~10).

\begin{algorithm}[!t]
  \caption{Linear Dynamic Programming (LDP)}
  \label{alg:Linear-dp}
    {\small
  \begin{algorithmic}[1]
    \STATE {\bfseries Input:} A linear computation graph $\mathcal{G'}$ and its size $n$
    \STATE {\bfseries Output:} All parallelization strategies in the cost frontier of execution time and memory consumption
    \STATE $\mathcal{CF}(o_1, s^k_1) = \mathcal{F}(o_1, s^k_1)$ for $s^k_1 \in \mathcal{S}_1$
    \FOR{\texttt{$i = 2$ to $n$}}
    \FOR{\texttt{$s^p_i \in \mathcal{S}_i$}}
    \STATE $\mathcal{CF}(o_i, s^p_i) = \cup_{s_{i-1}^k \in \mathcal{S}_{i-1} }\{\mathcal{F}(e_{(i-1)i},s_{i-1}^k,s_i^p)\otimes \mathcal{CF}(o_{i-1},s_{i-1}^k)\otimes \mathcal{F}(o_i,s_i^p)$\}
    \ENDFOR
    \ENDFOR
    \STATE $\mathcal{F}_o = reduce(\cup_{s_n^k\in\mathcal{S}_n}\mathcal{CF}(n,s_n^k))$
    \STATE Return $\mathcal{F}_o$
  \end{algorithmic}
}
\end{algorithm}

We denote the method that uses LDP to solve the cost frontier as \textit{FT-LDP} to contrast with FT-elimination. As we will show in Section~\ref{subsec:complexity analysis}, for a linear graph $\mathcal{G}'$ with $n$ operators and each operator has at most $K$ feasible parallelization configurations, the complexity of FT-LDP in Algorithm~\ref{alg:Linear-dp} is $O(n^2K^2\log(K)(\log(n)+\log(K)))$. In contrast, using FT-Elimination to track the cost frontier has a complexity of $O(n^2K^3\log(K)(\log(n)+\log(K)))$, which is much more costly than FT-LDP due to the large value of $K$. We will also show in the experiments that FT-LDP has much shorter running time than FT-elimination.

\vspace{1mm}
\textbf{Unroll LDP and elimination.} FT unrolls the strategy tuples in the final cost frontier $\mathcal{F}_o$ produced by LDP in Algorithm~\ref{alg:Linear-dp} to reconstruct the parallelization strategies for the entire computation graph $\mathcal{G}$. To provide information for unrolling, in each step of LDP and for each strategy tuple in $\mathcal{CF}(o_i, s^p_i)$, FT records the parallelization configuration of $o_{i-1}$ (i.e., $s_{i-1}^k$) and the strategy tuple in $\mathcal{CF}(o_{i-1},s_{i-1}^k)$ that produce it. Therefore, the final strategy tuples are unrolled by tracing back each step of LDP recursively. For unrolling elimination, FT records the parallelization configuration taken by the eliminated operator for each tuple in the cost frontier $\mathcal{F}$ produced by the elimination. Once we know the selected partial strategy in $\mathcal{F}$, the parallelization configuration of the eliminated operator can be reconstructed.

\vspace{1mm}
\textbf{Multi-threading for efficiency.} FT can be easily parallelized by multi-threading. For LDP, computing $\mathcal{CF}(o_i, s^p_i)$ for different parallelization configurations of operator $o_i$ (i.e., $s^p_i$) can be conducted in parallel as these computations only read $\mathcal{CF}(o_{i-1},s_{i-1}^k)$. Similarly, for the eliminations, the frontier updates for different parallelization configuration choices are also independent. For example, in node elimination, $\mathcal{F}(e_{hj}, s_h^w, s_j^p)$ under different $s_h^w$ and $s_j^p$ can be solved in parallel. Therefore, we spawn multiple threads to accelerate LDP and the eliminations.

\vspace{1mm}
\textbf{Improving cost estimation accuracy.} The memory consumption and execution time of the operators are relatively easy to predict~\cite{OptCNN,FlexFlow}. Thus the accuracy of cost estimation strongly depends on the quality of communication time (i.e., $t_x(e_{ij}, s_i^k, s_j^p)$ and $t_s(o_i, s_i^k)$) estimation. FlexFlow and OptCNN calculate the communication time using the amount of data to be transferred divided by the speed of the network connection between the devices. We found that this estimation method can lead to very large error (e.g., more than 70\%) for two main reasons. First, latency could dominate the communication time when transferring small tensors. Second, several communication operations could be executed by different devices simultaneously and these operations will contend for the PCIE or IB bandwidth, which makes communication time difficult to estimate.

We use collective operations for all the network communication and adopt a profile based method to estimate the commutation time. For collective communication operations, a parallel configuration of an operator divides the devices into disjoint groups (called device partitioning) and each group has the same amount of data to transfer. Although there is no communication between groups, different groups may still contend for bandwidth. Therefore, we profile the \textit{actual bandwidth} under different device partitioning schemes and data sizes. Specifically, under each device partitioning scheme, we measure the actual bandwidth for collective communication with a data size of $2^i$, in which $ 0 \!\leq \!i\! \leq \!P$ and $P$ is sufficiently large to cover all possible data sizes. When predicting the communication time for data with a size of $k$, we find the integer $i$ satisfying $2 ^i \leq k < 2 ^ {i + 1}$ and use the interpolation of the actual bandwidths at $2 ^ i$ and $2 ^ {i + 1}$. Our measurement shows that this method has an error of only $6\%-7\%$ in communication time estimation.

\subsection{Complexity Analysis}\label{subsec:complexity analysis}

In this part, we analyze the complexity of FT-LDP in Algorithm~\ref{alg:Linear-dp}. The results show that FT-LDP has a complexity that is quadratic in terms of the number of operators in the computation graph.


\begin{frontier complexity}\label{lemma:frontier complexity}
For a set $\mathcal{C}$ containing $K$ parallelization strategy tuples, its cost frontier can be obtained with a complexity of $O(K\log(K))$ using Algorithm~\ref{alg:frontier}.
\end{frontier complexity}

The proof of Lemma~\ref{lemma:frontier complexity} is straightforward as the complexity of Algorithm~\ref{alg:frontier} is dominated by sorting the $K$ tuples.

\begin{random tuple}\label{assup:random tuple}
For a set $\mathcal{C}$ containing $K$ strategy tuples, let $r_m(S_p, m_p, t_p)$ and $r_t(S_p, m_p, t_p)$ be the rank of tuple $(S_p, m_p, t_p)$ when sorting $\mathcal{C}$ in ascending order of memory and time consumption, respectively. $\mathcal{C}$ is said to have random order if $\mathcal{P}[r_m(S_p, m_p, t_p)=k]=\frac{1}{K}$ and $\mathcal{P}[r_t(S_p, m_p, t_p)=h]=\frac{1}{K}$ for $1 \le p, k, h \le K$, and $r_m(S_p, m_p, t_p)$ and $r_t(S_p, m_p,\\ t_p)$ are independent.
\end{random tuple}

In the following analysis, we always assume that a set $\mathcal{C}$ has \textit{random order} when solving its cost frontier. As we will see soon, the random order assumption implies that the cost frontier of a large set only has a small cardinality, which matches practice as most of the parallelization strategies are not favorable (i.e., both execution time and memory consumption are large).    

\begin{random size}\label{lemma:random size}
For tuple set $\mathcal{C}$ having random order and containing $K$ tuples, the expected size of its cost frontier $\mathcal{F}$ is $\log(K)$.
\end{random size}

\begin{proof}
  Denote the expected size of $\mathcal{F}$ as $f(K)$, where $K$ is the cardinality of the tuple set $\mathcal{C}$. Consider the tuple having the minimum time consumption in $\mathcal{C}$ (denoted as $S_p$), it is obvious that $S_p\in \mathcal{F}$ and tuples having larger memory consumption than $S_p$ do not belong to $\mathcal{F}$. The cost frontier of the tuples having smaller memory consumption than $S_p$ also belongs to $\mathcal{F}$, and the number of these tuples follows a discrete uniform distribution on $\{0,1,\cdots, K-1\}$ due to the random order assumption. Therefore, we can get the following recursive function,
  $$ f(K) = \sum_{k=1}^{K} \frac{f(k-1)}{K}  + 1.$$
  Solving the recursion gives $f(K) = \sum_{k=1}^{K} \frac{1}{k} = O(\log(K))$.
\end{proof}

We analyze the complexity of FT-LDP and FT-Elimination for frontier tracking \textit{when the computation graph $\mathcal{G}$ is a linear graph}. In this case, both $\mathcal{F}(e_{(i-1)i},s_{i-1}^k,s_i^p)$ and $\mathcal{F}(o_i,s_i^p)$ has a cardinality of 1. For more complicated graphs, the cardinalities of $\mathcal{F}(e_{(i-1)i},s_{i-1}^k,s_i^p)$ and $\mathcal{F}(o_i,s_i^p)$ depend on the elimination operations, which in turn depends on the specific structure of the computation graph. However, we also give the one-step complexity of FT-LDP and FT-Elimination when the cardinality of $\mathcal{F}(e_{(i-1)i},s_{i-1}^k,s_i^p)$ is not 1.


\begin{one step complexity}\label{theorem:one step complexity}
For FT-LDP in Algorithm~\ref{alg:Linear-dp}, assume that operators $o_{i-1}$ and $o_i$ both have $K$ parallelization configurations, the cumulative frontier $\mathcal{CF}(o_{i-1},s_{i-1}^k)$ of $o_{i-1}$  has a cardinality of $a$, and the edge cost frontier $\mathcal{F}(e_{(i-1)i},s_{i-1}^k,s_i^p)$ has a cardinality of $b$. The complexity of solving the cumulative frontier for $o_i$ (i.e., $\mathcal{CF} (i,s_i^p)$ for all $s_i^p\in\mathcal{S}_i$) is $O(K^2ab\log(Kab))$.
\end{one step complexity}


\begin{proof}
  According to the assumptions, $\mathcal{F}(e_{(i-1)i},s_{i-1}^k,s_i^p)\\\otimes \mathcal{CF}(o_{i-1},s_{i-1}^k)\otimes \mathcal{F}(o_i,s_i^p)$ has a cardinality of $ab$ as there is only one tuple in $\mathcal{F}(o_i,s_i^p)$ . $\cup_{s_{i-1}^k \in \mathcal{S}_{i-1} }\{\mathcal{F}(e_{(i-1)i},s_{i-1}^k,s_i^p)\otimes \mathcal{GF}(o_{i-1},s_{i-1}^k)\otimes \mathcal{F}(o_i,s_i^p)\}$ needs to enumerate all $K$ parallelization configurations of $o_{i-1}$ and thus has a cardinality of $Kab$. According to Lemma~\ref{lemma:frontier complexity} and Lemma~\ref{lemma:random size}, the cost frontier of $\mathcal{CF}(o_i, s^p_i)$ has an expected size of $O(\log(Kab))$ and finding it requires a complexity of $O(Kab\log(Kab))$. As all $K$ parallelization configurations of $o_i$ needs to be enumerated to find the cumulative frontier, the overall complexity is $O(K^2ab\log(Kab))$.
\end{proof}

\begin{linear dp complexity}\label{theorem:overall complexity}
For a linear computation graph $\mathcal{G}$ containing $n$ operators, and assume that each operator has at most $K$ parallelization configurations, the overall complexity of FT-LDP in Algorithm~\ref{alg:Linear-dp} is $O(n^2K^2\log(K)(\log(n)+\log(K)))$.
\end{linear dp complexity}

\begin{proof}
  For a linear graph $\mathcal{G}$, the cardinality of $\mathcal{CF}(o_1,s_1^k)$ is 1 for $1 \le k \le K$. The cardinality of $\mathcal{F}(e_{(i-1)i},s_{i-1}^k,s_i^p)$ is also 1 for any edge $e_{(i-1)i}$, $s_{i-1}^k$ and $s_i^p$. The expected cardinality of $\mathcal{CF}(o_{i-1},s_{i-1}^k)$ is bound by $O(\log(K^{i-2}))$ because there are $K^{i-2}$ partial parallelization strategies from operator $o_1$ to $o_{i-2}$ as each operator has $K$ parallelization configurations. According to Lemma~\ref{theorem:one step complexity}, the complexity for computing the cumulative frontier for operator $o_i$ is $O(K^2\log(K^{i-2})\log(K\log(K^{i-2})))$. Summing up the complexity from $o_2$ to $o_n$, we obtain the overall complexity of Algorithm~\ref{alg:Linear-dp} as $\sum_{i=2}^n O(K^2\log(K^{i-2})\log(K\log(K^{i-2})))$, which can be simplified as $O(n^2K^2\log(K)(\log(n)+\log(K)))$.
\end{proof}

For a linear computation graph $\mathcal{G}$, FT-Elimination conducts node elimination (as in Eq.~\ref{equ:node}) to simplify it to a graph that contains only two nodes. In the following, we analyze the complexity of node elimination and FT-Elimination.

\begin{node elimination complexity}\label{lemma:node elimination complexity}
For node elimination in Eq.~\ref{equ:node}, assume that the operators (i.e., $o_h$, $o_i$ and $o_j$) all have at most $K$ parallelization configurations, $\mathcal{F}(e_{hi}, s_h^w, s_i^k)$ and $\mathcal{F}(e_{ij}, s_i^k, s_j^p)$ have a cardinality of $a$ and $b$, respectively. Then node elimination has a complexity of $O(K^3ab\log(Kab))$.
\end{node elimination complexity}

\begin{proof}
  According to the assumptions, $\cup_{s_i^k\in\mathcal{S}_i}\{\mathcal{F}(e_{hi}, s_h^w, s_i^k) \\ \otimes \mathcal{F}(o_i, s_i^k)\otimes \mathcal{F}(e_{ij}, s_i^k, s_j^p)\}$ has a cardinality of $Kab$ and finding its cost frontier has a complexity of $O(Kab\log(Kab)$ according to Lemma~\ref{lemma:frontier complexity}. For node elimination, we need to enumerate the $K^2$ possible combinations of the possible parallelization configurations of operators $o_h$ and $o_j$ (i.e., $[s_h^w, s_j^p]$). Therefore, the overall complexity of node elimination is $O(K^3ab\log(Kab))$.
\end{proof}

\begin{optcnn complexity}\label{theorem:optcnn complexity}
For a linear computation graph $\mathcal{G}$ containing $n$ operators, and assume that each operator has at most $K$ feasible parallelization configurations, the overall complexity of using FT-Elimination for frontier tracking is $O(n^2K^3\log(K)\\(\log(n)+\log(K)))$.
\end{optcnn complexity}

\begin{proof}
  For a linear graph $\mathcal{G}$, both $\mathcal{F}(e_{hi}, s_h^w, s_i^k)$ and $\mathcal{F}(e_{ij}, s_i^k, s_j^p)$ have a cardinality of 1 initially. FT-Elimination will eliminate the nodes in $\mathcal{G}$ according to the topological order and each time it will eliminate the second node in the remaining graph. For the $i\textsuperscript{th}$ time of node elimination, $\mathcal{F}(e_{hi}, s_h^w, s_i^k)$ has a cardinality of $O(\log(K^{i-1}))$ while the cardinality of  $\mathcal{F}(e_{ij}, s_i^k, s_j^p)$ is 1. According to Lemma~\ref{lemma:node elimination complexity}, the $i\textsuperscript{th}$ node elimination has a complexity of $O(K^3\log(K^{i-1})\log(K\log(K^{i-1})))$. Summing up the complexity from $1$ to $n-2$, the result is $\sum_{i=1}^{n-2} K^3\log(K^{i-1})\log(K\\\log(K^{i-1}))$, which can be reduced to $O(n^2K^3\log(K)(\log(n)+\log(K)))$.
\end{proof}

Combining Theorem~\ref{theorem:overall complexity} and Theorem~\ref{theorem:optcnn complexity}, we can see that FT-LDP reduces the complexity of FT-Elimination by $K$ times when used for frontier tracking. For more complicated graphs, FT-LDP also has lower complexity than FT-Elimination as shown by the one-step complexity results in Lemma~\ref{theorem:one step complexity} and Lemma~\ref{lemma:node elimination complexity}.  

%% file: TO_system.tex
\section{The TensorOpt System}

MeshTensorFlow requires users to find the proper parallelization strategy by themselves and explicitly program the strategy. FlexFlow and OptCNN are based on Legion, which is not a popular system and does not have rich packages as in popular DL systems such as TensorFlow and PyTorch. ToFu is not open source and thus its usability remains unclear. Moreover, these systems cannot track the trade-off between different costs, which is important for scenarios such as scheduling in a multi-tenant cluster and price consideration on the could. To solve these problems, we develop a system called \textbf{TensorOpt} to make auto-parallelism user-friendly.

\subsection{Overall Description and API}

TensorOpt is built on top of TensorFlow, with a minimal extension of TensorFlow's API. TensorFlow scripts can be run as auto-parallel jobs on TensorOpt with only a few changes. Users only need to specify their preferences for parallelization strategy with some configurable options (to be introduced latter) and TensorOpt will invoke the FT algorithm to search for the desired parallelization strategy. TensorOpt also runs the chosen parallelization strategy automatically without user intervention and the details of parallel execution, e.g., the split of tensors among the GPUs and the communication among GPUs, are made transparent to users. 

When running DNN training jobs, several factors, e.g., efficiency, parallelism\footnote{Parallelism refers to the number of GPUs to be used, which also determines the amount of available memory and is important to the training throughput (i.e., the average number of training samples processed per second).} and cost, need to be considered. For a user who runs his job on an exclusive cluster, he may want to use all the GPUs in the cluster to minimize the execution time. But if the job is run on a shared cluster, the cluster scheduler may want to know the performance (i.e., training throughput) of the job under different parallelism to determine how much resource to allocate to run the job~\cite{gandiva}. When a user runs his job on the cloud, he may want to balance between cost and efficiency. Considering the different needs, TensorOpt currently provides the following three options for parallelization strategy search.  

\vspace{1mm}
\textbf{Mini-time} finds the parallelization strategy that minimizes the per-iteration time while satisfying the memory constraint, under a user-specified parallelism. This option is suitable for running jobs on pre-allocated devices or an exclusive cluster.

\vspace{1mm}
\textbf{Mini-parallelism} finds a parallelization strategy that requires the minimum number of devices (to satisfy the memory constraint). It may be used for program correctness checking or cost minimization. This is because per-GPU throughput usually decreases with parallelism and thus training with minimum parallelism is most cost effective. 

\vspace{1mm}
\textbf{Profiling} generates the minimum per-iteration time under a range of parallelism (without actually running the job), which is achieved by running the FT algorithm to minimize per-iteration time under these parallelisms. Note that a job may not be able to run if the parallelism is too small due to insufficient memory. If the parallelism is too large, per-iteration time may increase due to costly communication. This option can be used by the cluster scheduler or the cloud user to determine the proper parallelism for a job. Once the parallelism is determined, users can run TensorOpt in the \textit{mini-time} mode.  

%

%
%
%
%

\lstloadlanguages{Python}
\lstset{frame=tb,
  language=Python,
  aboveskip=3mm,
  belowskip=3mm,
  showstringspaces=false,
  columns=flexible,
  basicstyle={\small\ttfamily},
  numbers=left,
  numberstyle=\tiny\color{gray},
  keywordstyle=\color{blue},
  commentstyle=\color{dkgreen},
  stringstyle=\color{mauve},
  breaklines=true,
  breakatwhitespace=true,
  escapeinside={(*@}{@*)},
  tabsize=3
}

\begin{lstlisting}[language=Python, label={list:code}, caption=An example of using TensorOpt, captionpos=b, float=tp]
def create_model(input, labels):
  w = tensoropt.get_variable(...)
  b = tensoropt.get_variable(...)
  logits = tensoropt.matmul(input, w) + b
  return tensoropt.softmax_cross_entropy(labels, logits)

def training(loss):
  optimizer = tensoropt.GradientDescentOptimizer()
  train_op = optimizer.minimize(loss)
return train_op

def main():
  tensoropt.init()
  input = tensoropt.placeholder(...)
  label = tensoropt.placeholder(...)
  loss = create_model(input, label)
  train_op = training(loss)
  plan = tensoropt.find_strategy(option='mini_time')
  tensoropt.build_execution_graph(engine='tensorflow', plan=plan)
  tensoropt.run(train_op)
\end{lstlisting}

We provide an example script of using TensorOpt for DNN training in Listing~\ref{list:code}. The TensorOpt script is very similar to a TensorFlow script and there are only a few differences. We explain the key differences as follows.  
\begin{itemize}
  \item \textbf{init.} As TensorOpt uses the MPI library, the MPI environment is initialized at the beginning. Hardware and network information are also collected for use in parallelization strategy search during initialization.
        
  \item \textbf{find\_strategy.} Users provide their preferences for parallelization strategy with the aforementioned options and  the FT algorithm is invoked to find the suitable parallelization strategy according to user configuration.
        
  \item \textbf{build\_execution\_graph.} The execution graph for the low-level TensorFlow execution engine is constructed using the chosen parallelization strategy.
\end{itemize}

\subsection{System Design and Implementation}

\textbf{System workflow.}  Users define a computation graph using the high-level API and TensorOpt invokes the FT algorithm to find a proper parallelization strategy according to user configuration. Then TensorOpt spawns multiple processes (one for each device) and creates a TensorFlow execution graph for each processes according to the parallelization strategy. The execution graph describes how the job runs on multiple processes. We implemented wrappers for most of the key modules in TensorFlow, e.g., \textit{operator}, \textit{session}, and \textit{optimizer}. When creating the execution graph, TensorOpt propagates most of the parameters in the high-level API to the low-level API, e.g., the name of a tensor or operator, the initializer of a variable, and the strides or padding parameter for a convolution operator. However, the shapes of the tensors are not propagated to the execution graph as they are determined by the parallelization strategy. Users can use the distributed optimizers in TensorOpt in the same way as in TensorFlow and do not need to consider the details of parallelization.

TensorOpt also inserts communication operators into the execution graph for necessary communication among the processes. TensorOpt uses collective operations (e.g, \textit{allreduce} and \textit{allgather}) for all inter-device communication. Collective operations are more efficient and tractable (i.e., easy to predict performance) than peer to peer communication. A TensorOpt operator is decomposed into several TensorFlow primitive operators according to the need. For example, if the results need to be merged for matrix multiplication (e.g., $\mathbf{Y}=\mathbf{WX}$, with the model parameter $\mathbf{W}$ split along the column dimension), \textit{allreduce} is conducted after the TensorFlow matrix multiplication on each device.

\begin{figure}[!t]
  \centering
  \includegraphics[width=0.4\columnwidth]{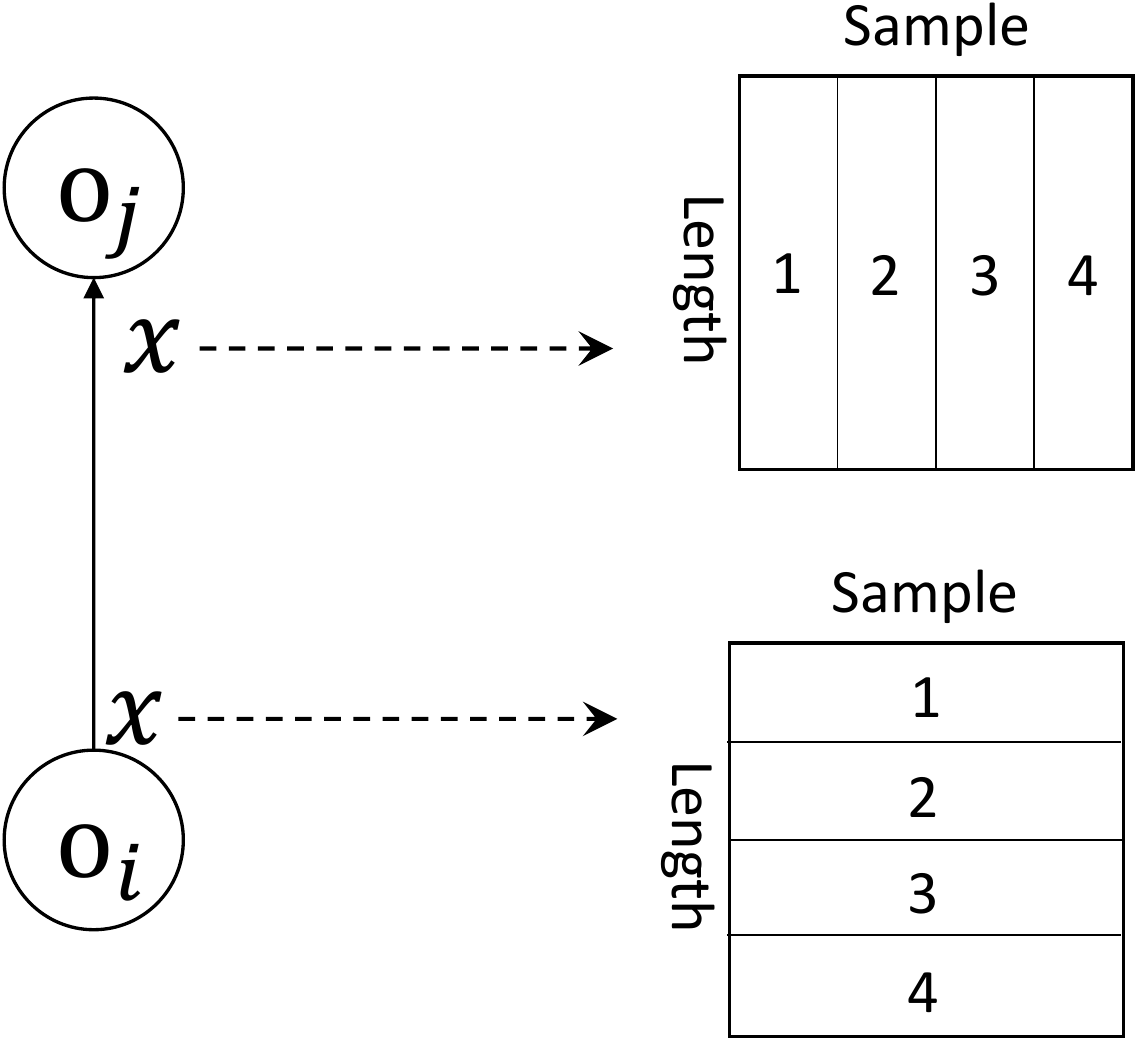}
  \caption{Example of tensor rescheduling, numbers 1-4 represent the splitting of the same tensor across the GPUs}
  \label{fig:reschedule}
\vspace{-3mm}
\end{figure}

\vspace{1mm}
\textbf{Flexible tensor splitting.}  MeshTensorFlow names each dimension of a tensor (called a logical dimension) and has two restrictions for splitting a tensor among the GPUs. First, the same device mesh is used for all operators in the computation graph. For example, four devices cannot switch between a one-dimensional mesh (i.e., $[4]$) and a two-dimensional mesh (i.e., $[2,2]$) for different operators. Second, if a logical tensor dimension is split across a device mesh dimension, then all operators having this tensor dimension also need to split across the device mesh dimension. For example, in a convolution neural network, if the batch dimension of the data tensor is split across all devices for the convolution layers (i.e., data parallelism), then the fully connected layers also need to be split in the batch dimension. However, model parallelism is usually more efficient for fully connected layers~\cite{OneWierdTrick}.

Obviously, the restrictions in MeshTensorFlow reduce the flexibility of parallelization strategies, and hence degrade the performance as we will show in Section~\ref{sec:anylysis}. Therefore, TensorOpt removes the two restrictions and allows different operators to have independent device mesh and tensor splitting. However, these flexibilities result in the \textit{re-scheduling problem} and we provide such an example in Figure~\ref{fig:reschedule}. Tensor $x$ splits among 4 GPUs in the \textit{length} dimension when generated as the output of operator $o_i$ but the downstream operator $o_j$ requires $x$ to split in the \textit{sample} dimension when used as input.

\begin{figure*}[!t]
	\centering
	\includegraphics[width=1.6\columnwidth]{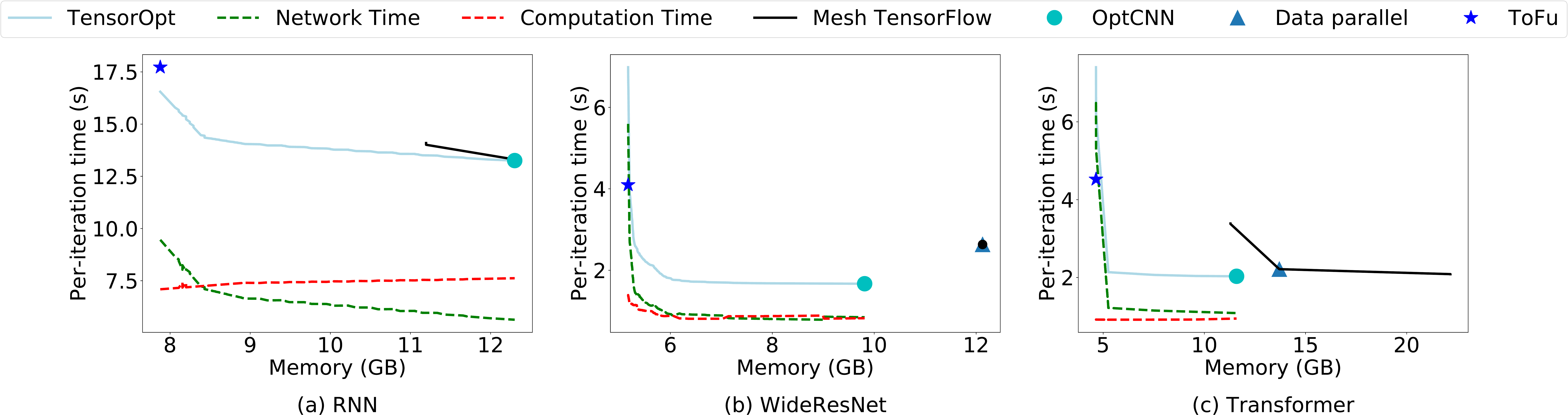}
	\vspace{-1.5mm}
	\caption{The cost frontier between memory consumption and execution time for some popular models, the solid lines are the cost frontiers while the dotted lines are the network time and computation time of TensorOpt (best viewed in color)}
	\label{fig:frontier}
	\vspace{-2mm}	
\end{figure*}

In this case, TensorOpt conducts~\textit{tensor re-scheduling} to adjust the output split of a tensor to the required input split.  Collective communication is used for tensor re-scheduling and TensorOpt finds the optimal communication operations by solving a shortest path problem. TensorOpt builds a graph, in which nodes are different tensor splits while an directed edge connects two tensor splits if one can be transformed into another with only one communication operation and the edge weight is the time taken by the communication. Thus, the optimal communication operations correspond to the shortest path from the output tensor split to the required input tensor split. TensorOpt fuses the sequence of communication operations into one operator to reduce intermediate memory usage. The FT algorithm also takes the cost of tensor re-scheduling into consideration (as edge cost) when tracking the cost frontier.                     

\vspace{1mm}
\textbf{Data loading.}  Existing auto-parallel systems (e.g., OptCNN and FlexFlow) require users to manually organize the data samples into the input split required by the parallelization strategy. For better usability, TensorOpt allows users to load training data by data parallelism and enjoy the data loading pipeline optimizations in popular deep learning frameworks such as TensorFlow and PyTorch. In this case, the operator that loads data is constrained to use data parallelism and the input data is adjusted to the desired input split via tensor re-scheduling when necessary. The cost of this re-scheduling is also considered when searching the parallelization strategy.

\vspace{1mm}
\textbf{Tensor reuse.} For some  tensors, both the output operator that generates them and the input operator that consumes them need them for backward propagation. For a tensor that needs re-scheduling, the two copies before and after re-scheduling are physically different (having different splits), and a straightforward solution is to keep both copies (i.e., one for the output operator and the other for the input operator). To save memory, TensorOpt allows \textit{tensor reuse} by providing three configurations for these tensors, i.e., keeping the copy before re-scheduling, keeping the copy after re-scheduling, and keeping both copies. If only one copy is kept, the other copy is reconstructed by re-scheduling when needed. Extra dependencies are inserted into the execution graph to ensure that tensor reuse is only activated during backward propagation. FT considers both memory and communication cost when choosing the configuration for a tensor. 


%% file: TO_analysis.tex
\begin{table}[t]
  \centering
  \caption{Statistics of the models}
  \label{tab:model}
  \vspace{-5.5mm}
  \begin{center}
    \scalebox{0.78}{\begin{tabular}{cccl}
        \toprule
        Model                          & Parameter (GB) & Batch Size & Memory (GB) \\
        \midrule
        RNN~\cite{lstm}                & 108            & 256        & 126         \\
        WideResNet~\cite{wideresnet}   & 7.3            & 256        & 83          \\
        Transformer~\cite{transformer} & 9.7            & 256        & 74          \\
        VGG16~\cite{VeryDeepCNN}       & 0.52           & 256        & 30          \\
        \bottomrule
      \end{tabular}}
  \end{center}
  \vspace{-5mm}
\end{table}

\section{Experimental Results}\label{sec:anylysis}

In our experimental evaluation, we first explore the trade-offs among different objectives (e.g., execution time, memory consumption, and network communication) for popular DNN models by analyzing their cost frontiers. Then we evaluate the accuracy and efficiency of the FT algorithm. We also test the efficiency of the TensorOpt system for distributed DNN training. In our cluster, each machine is equipped with 8 NVIDIA Tesla V100 GPUs (with 16 GB on chip memory), a 48-core Initel(R) Xeon(R) Platinum 8160 CPU and 256 GB main memory. The GPUs on the same machine use NVLink for communication, while GPUs on different machines use RDMA on 100 Gbps EDR Infiniband for communication. Unless otherwise stated, the experiments were conducted using 16 GPUs on two machines. The statistics of the models used in the experiments are listed in Table~\ref{tab:model}, where \textit{memory} is the \textit{estimated peak memory consumption} for training on a single GPU.

\subsection{Cost Frontier Analysis}

\begin{figure*}[!t]
  \centering
  \begin{subfigure}[b]{0.5\columnwidth}
    \includegraphics[width=\columnwidth]{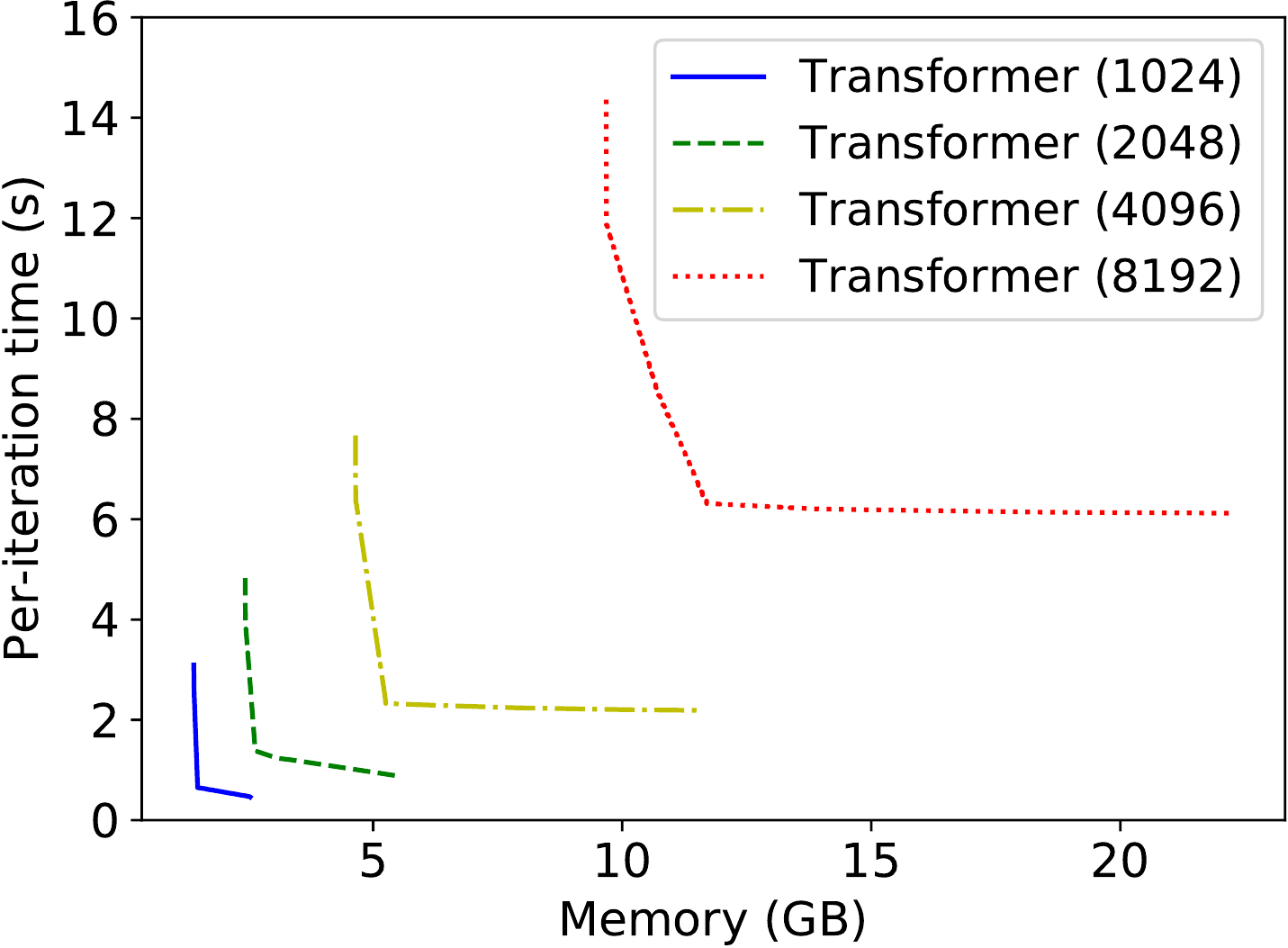}
    \caption{Model size}
    \label{fig:mode size}
  \end{subfigure}
  \begin{subfigure}[b]{0.5\columnwidth}
    \includegraphics[width=\columnwidth]{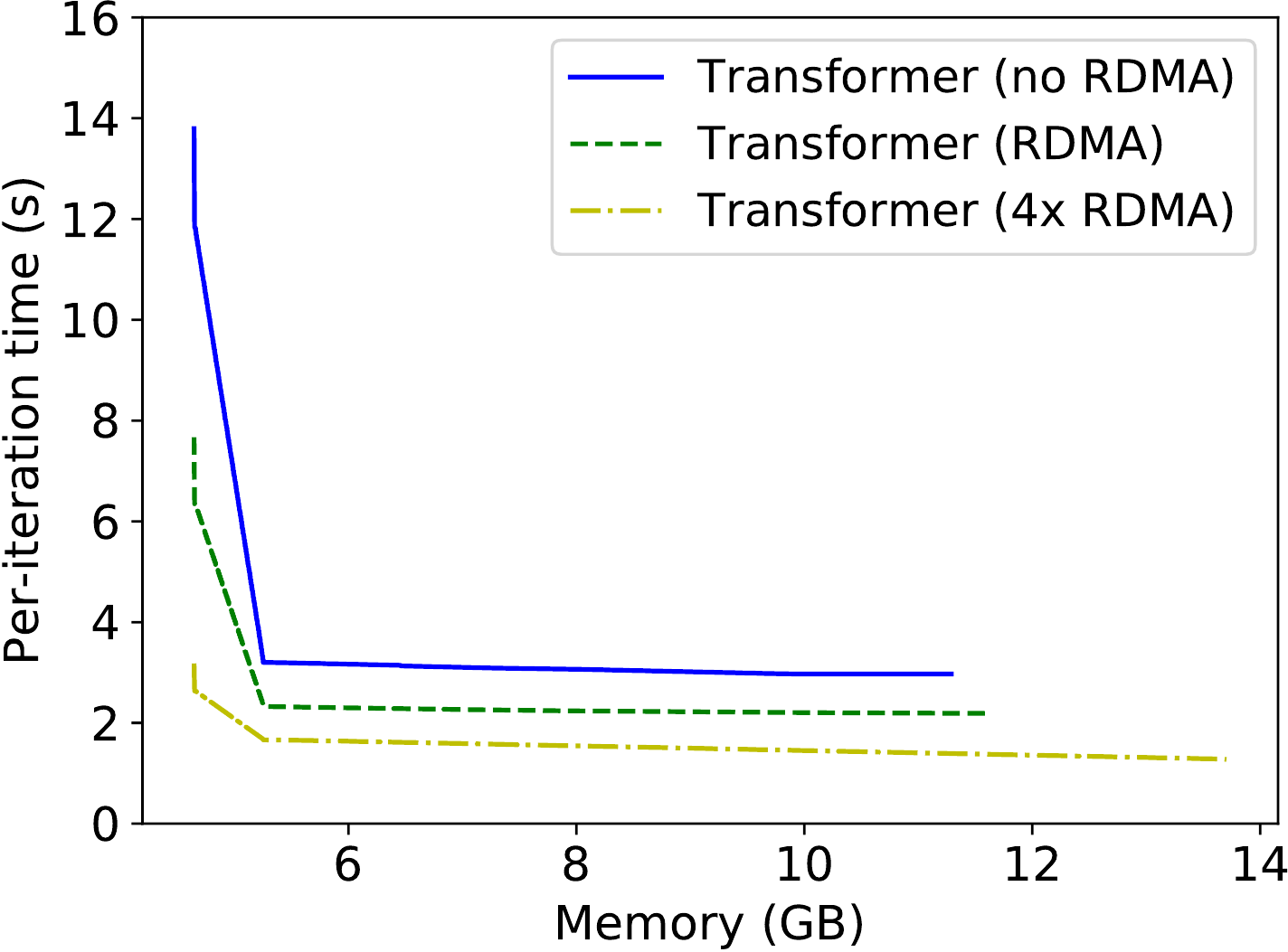}
    \caption{Cross-machine bandwidth}
    \label{fig:cross machine bandwidth}
  \end{subfigure}
  \begin{subfigure}[b]{0.5\columnwidth}
    \includegraphics[width=\columnwidth]{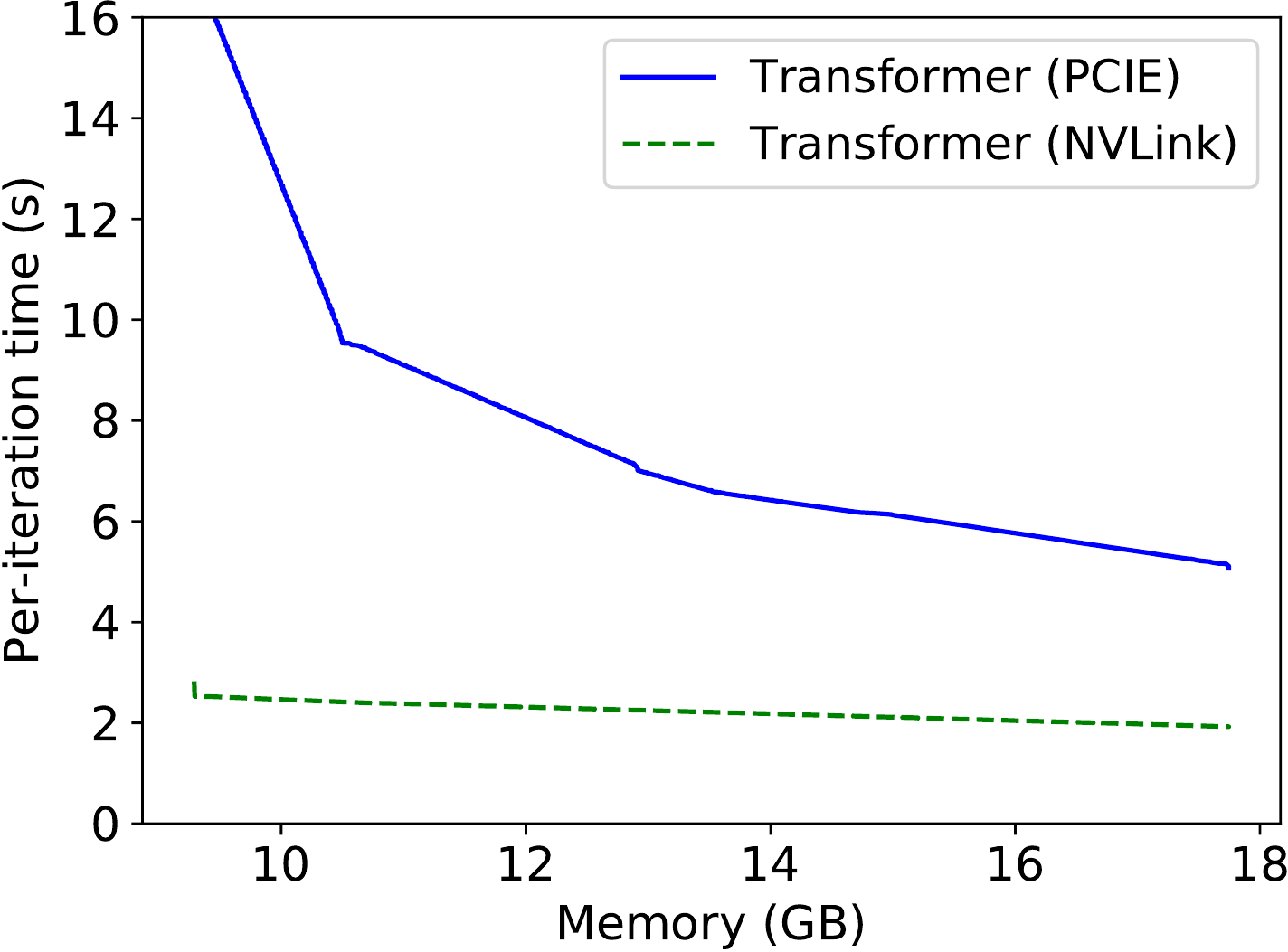}
    \caption{Intra-machine bandwidth}
    \label{fig:Intra machine bandwidth}
  \end{subfigure}
  \vspace{-2mm}
  \caption{The influence of different factors on the cost frontier for training Transformer using TensorOpt (best viewed in color)}
  \label{fig:factor}
    \vspace{-2mm}
\end{figure*}


\textbf{Cost frontier for different models.} In Figure~\ref{fig:frontier}, we plot the cost frontier between memory consumption and per-iteration time for some popular DNN models. Note that each point on the cost frontier represents a parallelization strategy and the coordinates of the point represent the memory consumption (on each GPU) and the per-iteration time of that strategy. We are interested in the large models as training large models is more challenging. The shape of the cost frontier for some small models (e.g., VGG16) is also similar to that of the large models (e.g., WideResNet). We also decompose the per-iteration time of TensorOpt into \textit{network time} and \textit{computation time}, and plot them using dotted lines. We did not include FlexFlow~\cite{Tofu} because both OptCNN~\cite{OptCNN} and FlexFlow optimize per-iteration time and they have similar performances for most of the workloads. We simulated ToFu using our cost model by splitting all the tensors among all the devices and disabling tensor replication. For Mesh-TensorFlow, we solved its cost frontier by adding the tensor split restrictions. Data Parallel, OptCNN and ToFu provide a single strategy instead of tracking the cost frontier, and thus each of them corresponds to only one point in Figure~\ref{fig:frontier}. For RNN, the performance of Data Parallel is poor (taking 109 GB memory and 39 seconds per iteration) and we do not plot it in the figure for clear presentation of the results of the other methods. For WideResNet, the cost frontier of MeshTensorFlow is a single point that collides with Data Parallel. From the results in Figure~\ref{fig:frontier}, we can make the following observations.

First, the computation time remains stable under different parallelization strategies for TensorOpt, but the network communication time decreases when using more memory, which causes the per-iteration time to decrease. Therefore, the dotted green line can also be regarded as the approximate cost frontier between network communication time and memory consumption. For WideResNet, the computation time increases when memory is limited, because the parallelization strategies conduct redundant computation on different GPUs to reduce network communicansun.

Second, for all three models, the network communication time (and hence the per-iteration time) drops rapidly when we increase available memory to a certain threshold and remains relatively stable when memory exceeds the threshold. We call the point at this threshold the \textit{turning point} on the cost frontier. We found that when memory is limited, tensors that need re-scheduling will keep only one copy and a re-scheduling is needed to reconstruct another copy during back propagation, which incurs communication overhead. When the amount of memory increases, the parallelization strategies tend to keep both copies for these tensors and thus the network communication time drops. It is difficult to further reduce the network communication time when memory is already sufficient as most re-scheduled tensors have enough space to keep both copies. From an economical point of view, the memory used at the turning point may be a suitable choice for memory provision as using less memory will significantly degrade the per-iteration time but investing more memory has only marginal performance benefits.

Third, by removing the restrictions on tensor split in MeshTensorFlow, TensorOpt significantly outperforms MeshTensorFlow. For both RNN and Transformer, the cost frontier of TensorOpt is always below that of MeshTensorFlow, meaning that TensorOpt has shorter per-iteration time using the same amount of memory. Moreover, MeshTensorFlow cannot work in the small-memory region, meaning that the minimum memory needed by MeshTensorFlow is significantly higher than that required by TensorOpt. For WideResNet, the optimal strategy of MeshTensorFlow is data parallel because the initial layers dominate the overall complexity and favor data parallel, and MeshTensorFlow cannot switch to other configurations for the other layers due to its restrictions.


Finally, for all three models, Data Parallel has poor performance with large memory consumption and long per-iteration time. OptCNN always finds the point with the shortest per-iteration time on TensorOpt's cost frontier as it is designed to minimize the per-iteration time. In contrast, ToFu always uses a small amount of memory with a long per-iteration time. Compared with OptCNN and ToFu, TensorOpt can work for any point on the frontier, which brings better flexibility to adapt to resource availability and cost-efficiency trade-offs.

\vspace{1mm}

\textbf{Influence of different factors.} To better understand the influence of different factors on the cost frontier, we plot the cost frontier for training Transformer using TensorOpt under different model sizes and network settings in Figure~\ref{fig:factor}. In Figure~\ref{fig:factor}~(a), we control the model size of Transformer by adjusting its hidden size. The results show that for the same model structure with different sizes, the cost frontiers have similar shape but the turning point has larger memory consumption for larger models. In Figure~\ref{fig:factor}~(b), \textit{no RDMA} uses Infiniband directly (by disabling RDMA) for cross-machine communication and the bandwidth becomes approximately 0.5 times of RDMA, while \textit{4x RDMA} assumes the cross-machine bandwidth is 4 times of RDMA and corresponds to NVIDIA DGX, which has 4 Infiniband network cards. The results show that the cost frontiers have similar shape and the memory consumptions at the turning point are almost identical for different configurations. This is because under all three cases, cross-machine communication is slower than intra-machine communication (e.g., even \textit{4x RDMA} is 10 times slower than NVLink) and thus the parallelization strategies will always try to reduce the amount of cross-machine communication. However, the per-iteration time of \textit{4x RDMA} is only half of \textit{no RDMA} at the turning point, which suggests that cross-machine bandwidth has a big impact on the performance. In Figure~\ref{fig:factor}~(c), we train the model with 8 GPUs on a single machine but use different methods for intra-machine communication. The bandwidth of PCIE is approximately 1/20 of NVLink according to our measurement. The results show that using NVLink provides a significant reduction in the per-iteration time compared with using PCIE at the same memory consumption.

From the results reported in Figures~\ref{fig:factor}~(a)-(c), different model sizes and network settings may result in different parallelization strategies with significantly different costs. As it is non-trivial to find the optimal parallelization strategy given a particular network setting (or other hardware setting) and model size, the ability to track the cost frontier makes the FT algorithm a powerful tool to efficiently characterize the influence of various factors on the training performance.



\begin{figure}[!t]
		\vspace{2mm}
	\centering
	\begin{subfigure}[b]{0.49\columnwidth}
		\includegraphics[width=\columnwidth]{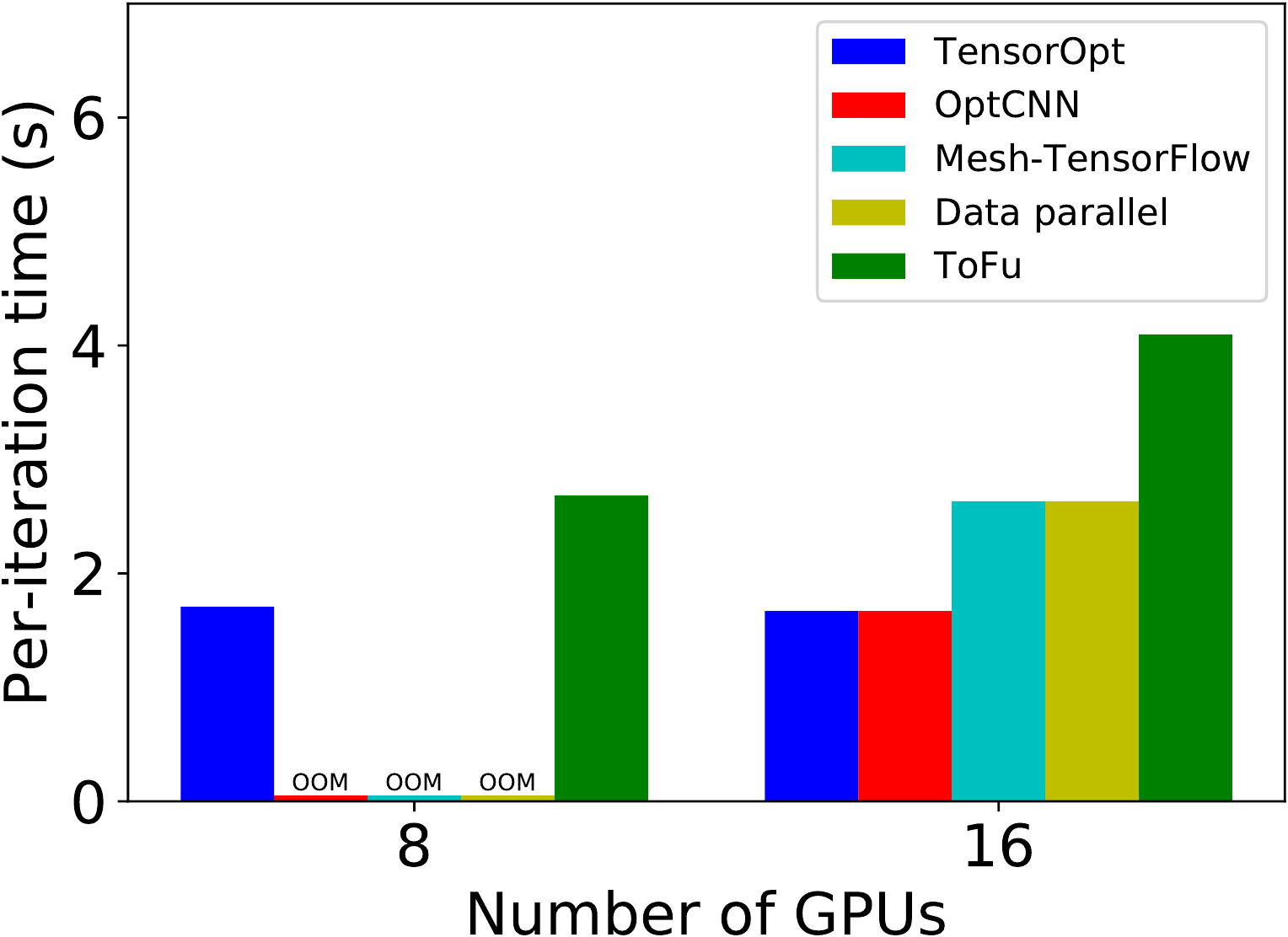}
		\caption{WideResNet}
		\label{fig:parallelism wrs}
	\end{subfigure}
	\begin{subfigure}[b]{0.49\columnwidth}
		\includegraphics[width=\columnwidth]{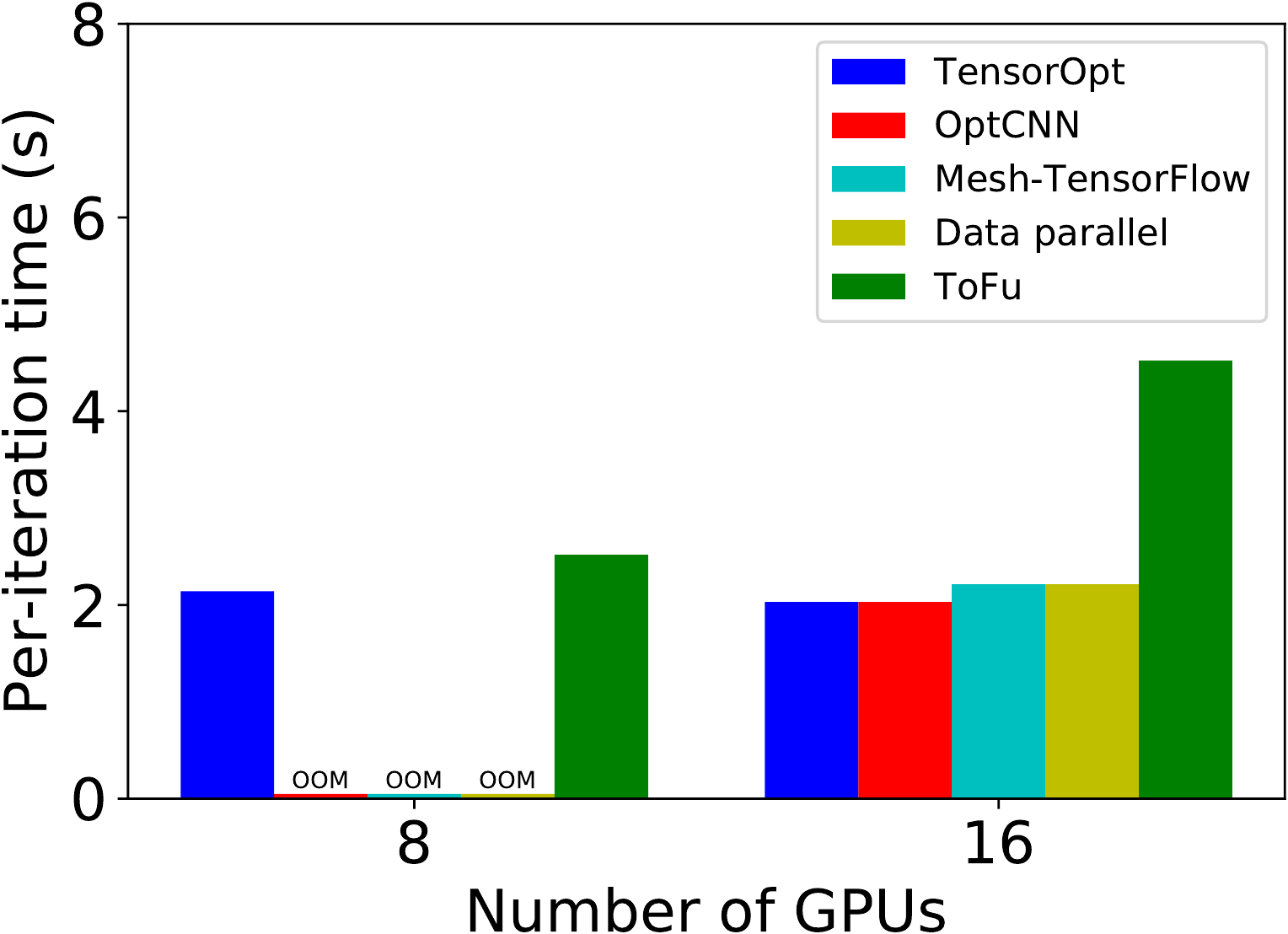}
		\caption{Transformer}
		\label{fig:parallelism Transformer}
	\end{subfigure}
	\vspace{-5mm}
	\caption{The relation between parallelism and the shortest per-iteration time for different models (best viewed in color)}
	\label{fig:parallelism}
	\vspace{-2mm}
\end{figure}

\vspace{1mm}

\textbf{Flexibility in adapting to resource availability.} One unique advantage of the FT algorithm over existing parallelization strategy search algorithms is its flexibility in adapting to different resource situations. We illustrate this phenomenon by plotting the relation between per-iteration time and parallelism for WideResNet and Transformer in Figure~\ref{fig:parallelism}. Note that in practice, we cannot change the on-chip memory of the GPUs, and the amount of memory is actually controlled by parallelism (i.e., providing more memory by using more GPUs). The results show that when the number of GPU is small (e.g., 8), Data Parallel and OptCNN cannot run the training job but TensorOpt can. For both models, running with 8 GPUs may be the most cost-effective because the per-iteration time only reduces marginally for TensorOpt when increasing to 16 GPUs (possibly because of expensive cross-machine communication). However, Data Parallel and OptCNN require at least 16 GPUs to run. ToFu can run under a small parallelism but the per-iteration time even increases when more GPUs are provided. We found this is because ToFu excessively minimizes memory consumption, which incurs a large amount of costly cross-machine communication when using 16 GPUs. TensorOpt is flexible in using different levels of available resources because it tracks the cost frontier and can select any strategy on the frontier according to resource availability. When the number of GPUs is small, TensorOpt chooses a strategy with low memory consumption. But TensorOpt can also minimize the per-iteration time when the number of GPUs is sufficient. Moreover, the strategy transition in TensorOpt is seamless and automatic with the cost frontier.

\subsection{Accuracy and Efficiency of FT}

We use the FT algorithm to track the cost frontier and estimate the costs of the parallelization strategies. Thus, it is important that the algorithm provides an accurate cost estimation and runs efficiently.

We report the cost estimation error of FT for different models in Table~\ref{tab:estimation error}. The error is defined as $(c-\hat{c})/c$, where $c$ is the actual cost and $\hat{c}$ is the estimated cost. The reported error is the average of 20 randomly sampled parallelization strategies for each model. The results show that FT has a small estimation error (below 8\% in all cases) and consistently underestimates the costs. We found that FT underestimates the network communication time (and hence the overall execution time) because some communication overheads are not taken into consideration, e.g., the progress synchronization among the devices and the coordination messages for collective communication. FT underestimates the memory consumption because there are some temporary tensors that take up memory. To prevent TensorOpt from running out of memory, we can choose a parallelization strategy that has slightly lower memory consumption than the devices on-chip memory. For example, for GPUs with 16GB memory, a parallelization strategy with 14.5GB ($\approx 16/1.1$) peak memory consumption would be safe. We also found that using the simplified method in OptCNN and FlexFlow for communication time estimation (i.e., dividing the data volume by the network bandwidth) leads to large errors in cost estimation. For example, its estimation error in the network communication time is 74.8\% for RNN.

\begin{table}[t]
  \centering
  \caption{Estimation error of the FT algorithm }
  \label{tab:estimation error}
    \vspace{-6mm}
  \begin{center}
    \scalebox{0.77}{\begin{tabular}{cccl}
      \toprule
      Model       & Execution Time & Network Time & Memory \\
      \midrule
      RNN         & 7.16\%         & 7.16\%       & 4.86\% \\
      WideResNet  & 7.62\%         & 3.05\%       & 4.47\% \\
      Transformer & 5.02\%         & 7.23\%       & 0.98\% \\
      \bottomrule
    \end{tabular}}
  \end{center}
    \vspace{-5.5mm}
\end{table}

We report the running time of the FT algorithm for different models when tracking the cost frontier under 16 GPUs in Table~\ref{tab:running time}. The results were measured using the CPU of a single machine in our cluster. FT-Elimination uses elimination to simplify the graph to only two nodes, while \textit{no multi-thread} disables the multi-threading in FT-LDP. The results show that FT-LDP has significantly shorter running time than FT-Elimination, which is consistent with the complexity analysis in Section~\ref{subsec:complexity analysis}. Multi-threading also effectively reduces the running time, especially for models with a large number of operators (e.g., WideResNet). Overall, the running time of FT-LDP is acceptable  (tens of minutes for very complex models) considering the long training time of DNN models (e.g., days or even weeks).

\begin{table}[t]
  \centering
  \caption{Running time of the FT algorithm (in seconds)}
  \label{tab:running time}
    \vspace{-6mm}
  \begin{center}
    \scalebox{0.75}{\begin{tabular}{cccl}
        \toprule
        Model                    & WideResNet & RNN  & Transformer \\
        \midrule
        FT-LDP                   & 1,292      & 0.28 & 201         \\
        FT-Elimination           & 19,666    & 1.78 & 3,030       \\
        FT-LDP (no multi-thread) & 17,432     & 0.40 & 1,535       \\
        \bottomrule
      \end{tabular}}
  \end{center}
    \vspace{-3mm}
\end{table}

\subsection{Efficiency of TensorOpt}

We evaluated the efficiency of TensorOpt by comparing with Horovod~\cite{horovod} for training different models with 16 GPUs. Horovod is the state-of-the-art execution engine for data parallelism. We did not compare with ToFu because it is not open source. We also did not compare with MeshTensorFlow because it is hard to tune the parallelism strategy to run since MeshTensorFlow can only set the strategy manually. As OptCNN and FlexFlow are based on Legion, the comparison may not be fair due to the differences in execution engine. Horovod uses data parallelism for training and (in a way similar to TensorOpt) delegates single machine execution to TensorFlow. We used two configurations for TensorOpt, \textit{mini-time} means minimizing the per-iteration time under the given parallelism, while \textit{data parallel} uses the same parallelization strategy as Horovod. The Transformer model used in this experiment (with 4.8GB parameter) is smaller than the one in Table~\ref{tab:model} as Horovod cannot run the large model.

The results in Table~\ref{tab:efficiency comparison} show that TensorOpt in the \textit{mini-time} mode achieves significantly shorter running time than Horovod for VGG16 and WideResNet, which validates the advantage of auto parallelism. In the \textit{data parallel} model, TensorOpt has slightly longer per-iteration time than Horovod and we found that this is because Horovod only considers data parallelism so that it can merge the synchronization for small tensors to fully utilize the bandwidth. However, in auto-parallelism, we cannot merge communication operations as some operations may block the computation. For Transformer, the three configurations have similar performances because the per-iteration time of data parallel is close to minimum.

\begin{table}[t]
  \centering
  \caption{Per-iteration time for TensorOpt and Horovod (s)}
  \label{tab:efficiency comparison}
    \vspace{-6mm}
  \begin{center}
    \scalebox{0.78}{\begin{tabular}{cccl}
        \toprule
        Model                     & VGG16 & WideResNet & Transformer-S \\
        \midrule
        TensorOpt (mini-time)     & 0.10  & 1.99       & 1.16          \\
        TensorOpt (data parallel) & 0.16  & 2.89       & 1.18          \\
        Horovod                   & 0.15  & 2.80       & 1.04          \\
        \bottomrule
      \end{tabular}}
  \end{center}
    \vspace{-4.5mm}
\end{table}

%% file: TO_conclusion.tex
\section{Conclusions}\label{sec:conclusion}  


We presented the FT algorithm for parallelization strategy search and the TensorOpt system for distributed DNN training. The flexibility of FT allows us to train large models with limited memory or maximize training efficiency when memory is sufficient. Based on FT, TensorOpt makes distributed DNN training more user-friendly by automatically searching and executing  parallelization strategies. Using TensorOpt is as easy as vanilla TensorFlow, and users only need to define the computation graph and provide the preference for parallelization strategy. Our experimental results validates the effectiveness of the FT algorithm for parallelization strategy search and the flexibility of TensorOpt in distributed DNN training given different resource availability.